\documentclass[sigconf]{acmart}
\newcommand{\sysname}{GenAssist}
\newcommand{\ipstart}[1]{\vspace{1mm}\noindent{\textbf{\textit{#1.}}}}
\usepackage{color}
\usepackage{stfloats}
\usepackage{colortbl}
\usepackage{multirow}

\newcommand{\revised}[1]{{\color{black}{ #1}}}

\definecolor{mina}{RGB}{148, 0, 211}
\definecolor{yihao}{RGB}{27,158,119}
\definecolor{amy}{RGB}{2,95,217}

\AtBeginDocument{%
  \providecommand\BibTeX{{%
    \normalfont B\kern-0.5em{\scshape i\kern-0.25em b}\kern-0.8em\TeX}}}

\setcopyright{acmcopyright}
\copyrightyear{2018}
\acmYear{2018}
\acmDOI{XXXXXXX.XXXXXXX}

\acmConference[Conference acronym 'XX]{Make sure to enter the correct
  conference title from your rights confirmation emai}{June 03--05,
  2018}{Woodstock, NY}
\acmBooktitle{Woodstock '18: ACM Symposium on Neural Gaze Detection,
 June 03--05, 2018, Woodstock, NY} 
\acmPrice{15.00}
\acmISBN{978-1-4503-XXXX-X/18/06}

\begin{document}

\title{GenAssist: Making Image Generation Accessible}



\author{Mina Huh}
\affiliation{
    \institution{The University of Texas at Austin}
    \city{Austin}
    \state{TX}
    \country{USA}
    }
\email{minahuh@cs.utexas.edu}

\author{Yi-Hao Peng}
\affiliation{
    \institution{Carnegie Mellon University}
    \city{Pittsburgh}
    \state{PA}
    \country{USA}
    }
\email{yihaop@cs.cmu.edu}

\author{Amy Pavel}
\affiliation{
    \institution{The University of Texas at Austin}
    \city{Austin}
    \state{TX}
    \country{USA}
    }
\email{apavel@cs.utexas.edu}

\begin{abstract}

Blind and low vision (BLV) creators use images to communicate with sighted audiences. However, creating or retrieving images is challenging for BLV creators as it is difficult to use authoring tools or assess image search results. Thus, creators limit the types of images they create or recruit sighted collaborators. 
While text-to-image generation models let creators generate high-fidelity images based on a text description (i.e. prompt), it is difficult to assess the content and quality of generated images. We present \sysname{}, a system to make text-to-image generation accessible. Using our interface, creators can verify whether generated image candidates followed the prompt, access additional details in the image not specified in the prompt, and skim a summary of similarities and differences between image candidates. 
To power the interface, \sysname{} uses a large language model to generate visual questions, vision-language models to extract answers, and a large language model to summarize the results. 
Our study with 12 BLV creators demonstrated that \sysname{} enables and simplifies the process of image selection and generation, making visual authoring more accessible to all.

\end{abstract}

\begin{teaserfigure}
  \centering
  \includegraphics[width=\textwidth]{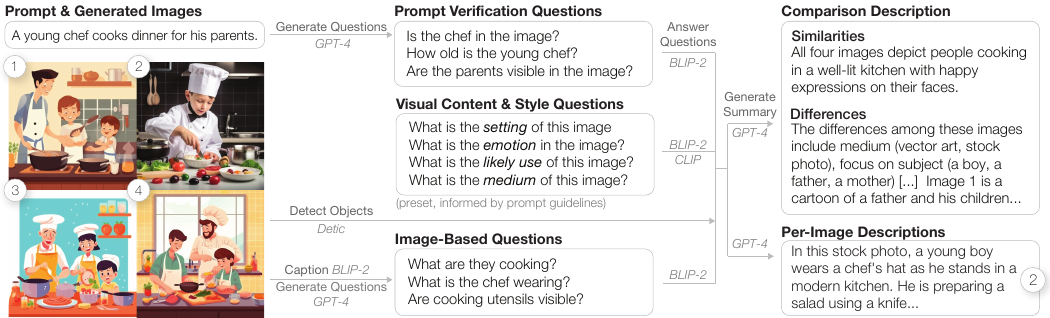}
  \caption{\sysname{} makes image generation accessible by providing rich visual descriptions of image generation results. Given a text prompt and set of generated images, \sysname{} uses a large language model (GPT-4) to generate \textit{prompt verification questions} from the prompt and \textit{image-based questions} from the image captions. \sysname{} then answers the visual questions (BLIP-2) and uses a vision-language model (CLIP) and an object detection model (Detic) to extract additional visual information. \sysname{} then uses GPT-4 to summarize all of the information into \textit{comparison descriptions} and \textit{per-image descriptions}.} 
  \label{fig:teaser}
  \Description{Teaser image that illustrates how \sysname{} generates the comparison description and per image descriptions in the summary table. First, \sysname{} takes the input of text prompt ``A young chef is cooking dinner for his parents'' and the four images generated using the prompt. Then, based on the prompt, \sysname{} uses GPT4 to ask prompt verification questions and use BLIP2 to answer them. \sysname{} also asks questions based on the individual image captions using GPT4 and BLIP2. In addition to prompt verification questions and image based questions, \sysname{} also asks questions related to visual content and styles and answer them using BLIP2, Detic and CLIP. Finally, using all of the visual information, the comparison description (similarities and differences) and the per image descriptions are generated using GPT4.}
\end{teaserfigure}

\maketitle

\section{Introduction}

BLV creators use images in presentations~\cite{peng2022diffscriber}, social media~\cite{bennett2018teens}, videos~\cite{huh2023avscript}, and art~\cite{bornschein2017digital}.
To obtain images, creators currently either describe their desired images to the sighted collaborators who then search for or create the image~\cite{peng2022diffscriber, wang2021revamp}, or limit the types of images they create~\cite{schaadhardt2021understanding}. Large-scale text-to-image generation models, such as DALL-E~\cite{ramesh2021zero}, Stable Diffusion~\cite{rombach2021highresolution}, and Midjourney~\cite{midjourney}, present an opportunity for these creators to generate images directly from text descriptions (i.e., prompts). However, current text-to-image generation tools are inaccessible to BLV creators, as creators must \textit{visually inspect} the content and quality of the generated images to iteratively refine their prompt and select from multiple generated candidate images. 

While BLV creators can gain access to images using automated descriptions~\cite{seeingAI,li2023blip}, existing descriptions are intended primarily for image consumption. As a result, the descriptions leave out details that may help authors decide whether or not to use the image (\textit{e.g.}, style, lighting, colors, objects, emotions).
Prior work also enables users to gain flexible access to the spatial layout of objects in images~\cite{lee2021image}, but exploring details per image makes it difficult to assess similarities and differences between image options provided during image generation. 
To make authoring visuals more accessible, prior work has explored describing visuals to help creators author presentations~\cite{peng2022diffscriber} or videos~\cite{huh2023avscript}. While such work helps creators identify low-quality visuals (\textit{e.g.}, blurry footage in a video~\cite{huh2023avscript}) or graphic design changes (\textit{e.g.}, changing slide layouts~\cite{peng2022diffscriber}), prior work has not yet explored how to improve the accessibility of image generation.


To understand the opportunities and challenges of text-to-image generation, we conducted a formative study with 8 BLV creators who regularly create or search for images.
Creators in our study reported their existing strategies for making images themselves (\textit{e.g.}, using SVG editors or code), searching for images, or asking others to search for or create images (similar to prior work~\cite{peng2022diffscriber,huh2023avscript,bennett2018teens}). All creators expressed excitement about using image generation to improve their efficiency and expressivity in image authoring. 
Creators all used image generation for the first time during our study and enjoyed creating high-fidelity images for their own uses (\textit{e.g.}, creating a logo for their website, making a card for their family). 
While we invited participants to ask the researchers visual questions to gain access to the visual details (\textit{e.g. ``What are the differences?'', ``Is the color calm or aggressive?''}), it remained challenging for participants to: craft a well-specified prompt especially without visual experience, assess how well the generated image followed the prompt, recognize generated details that were not originally specified in the prompt, and understand or remember the similarities and differences between images.

To improve the accessibility of image generation, we present \sysname, a system that provides access to text-to-image generation results via prompt-guided image descriptions and comparisons (Figure~\ref{fig:teaser}). 
Our system lets creators skim an overview of similarities and differences between images using our comparison descriptions and per image descriptions (Figure~\ref{fig:teaser}, right), assess if the images followed their prompt using prompt verification (Figure~\ref{fig:teaser}, center), and recognize visual details not in the prompt using our content and style extraction (Figure~\ref{fig:teaser}, center).
Creators can also interactively ask questions across multiple images to gain additional details.
Our interface design enables creators to easily navigate visual information via a screen reader-accessible table format. Our tables let creators selectively gain information about individual images (columns) or visual questions (rows) (Figure~\ref{interface}).


We evaluated \sysname{} in a within-subjects study with 12 BLV creators who compared \sysname{} with a baseline interface that was designed to encompass practices of accessing images (\textit{e.g.,} automated caption~\cite{wu2017automatic}, object detection~\cite{seeingAI}, and Visual Question Answering~\cite{li2023blip}). Participants rated \sysname{} as more useful than the baseline interface for understanding similarities and differences between the images, and they reported higher satisfaction with their image generation performance. 
Participants all expressed excitement about using \sysname{} in their own workflows for authoring images and for new uses.

Our work contributes:
\begin{itemize}
    \item Design opportunities making image generation accessible, derived from a formative study
    \item \sysname, a system that provides access to image generation results via prompt-guided summaries and descriptions
    \item User study that demonstrates how BLV creators use \sysname{} to interpret and generate images
\end{itemize}

\section{Background}

As we aim to enhance the experience of content BLV creators working with AI-powered image-generation tools, our work builds upon prior research that explores: the accessibility of authoring tools and images, and text-to-image generation tools.

\subsection{Accessibility of Authoring Tools}

Enabling access to authoring tools unlocks new forms of self-expression.
Recent research investigated how BLV people take and edit photos and videos~\cite{bennett2018teens, huh2023avscript}, compose music~\cite{payne2020blind}, draw digital images~\cite{bornschein2017digital}, and make presentations~\cite{schaadhardt2021understanding, peng2022diffscriber, zhang2023a11yboard}. Such work includes studies of current practices that highlight accessibility concerns of existing authoring tools and the authored visuals. For example, features of current authoring tools remain difficult to access using screen readers~\cite{li2021accessibility, huh2023avscript, peng2021say}, and it can be difficult to assess the effect of the visual edits such as color changes~\cite{schaadhardt2021understanding}.

To improve the accessibility of authoring tools, researchers have explored methods for providing feedback to authors as they modify visual elements. For example, prior work has developed tactile devices that assist BLV designers in understanding and adjusting the layout of user interface elements~\cite{li2019editing, potluri2019multi}. Tactile feedback has also been used to help developers interpret code structure, such as indentation~\cite{falase2019tactile}.
Other prior work has used audio notifications to inform users about scene changes when reviewing videos~\cite{huh2023avscript, peng2021slidecho}, while text descriptions have been used to convey visual details important to authoring such as brightness and layout~\cite{huh2023avscript, peng2022diffscriber}. Sound and text feedback have also been used to keep blind authors informed about their collaborators' edits to documents~\cite{lee2022collabally}.
Similar to prior research, we also aim to make authoring tools accessible by providing in-situ feedback, but we instead provide creation-specific information to facilitate authoring images.

In addition to offering authoring feedback, researchers have developed systems to automate visual authoring. Prior systems recommend 2D layouts for visual elements during graphic design~\cite{o2015designscape} and transform text into visual presentations~\cite{leake2020generating, xia2020crosspower, sefid2021slidegen}. To accommodate individual preferences and mitigate the impact of errors produced during generation, these systems typically offer multiple options for users to choose from and allow iterative generation attempts. Iterative generation and selection are not accessible for BLV creators, as it requires visually inspecting the output designs to choose a generated option or revise the input. In this work, we seek to make automated authoring tools, such as image generation, more accessible to BLV creators. Our approach provides a structured format for assessing and comparing generated results, and on-demand access to additional visual details to support creators in selecting a result and revising their input. 



\subsection{Accessibility of Images}

Improving the accessibility of image generation systems involves not only ensuring access to image generation features, but also making the produced images accessible. A primary method for making images more accessible is representing them as text descriptions, such as image captions or alt text (\textit{e.g., ``A person walking on the street''}). Early work hired crowd workers to create alt text~\cite{von2004labeling, bigham2010vizwiz}, while recent research has developed machine-learning-based systems that automatically generate image descriptions~\cite{xu2015show, vinyals2015show, li2023blip}. 
Building on auto-generated captions, researchers have developed systems that further improve users' understanding of images by providing additional information, such as regional descriptions~\cite{zhong2015regionspeak,seeingAI}, and structuring detailed descriptions into an overview~\cite{lee2021image, facebook-detailed-description, morris2018rich}. This approach enables users to review visual information more efficiently and has been found to help blind people better understand images compared to using captions alone~\cite{lee2022imageexplorer}. Our work builds upon this idea by presenting descriptions of image generation results in a hierarchical, easy-to-compare format, and tailoring the descriptions to the task of authoring rather than consuming images.


\revised{Automatic descriptions do not always capture all of the important image details. Visual Question Answering (VQA) tools can fill this gap by offering on-demand information to visual questions (\textit{e.g.}, \textit{``What is the person walking on the street wearing?''}). 
Previous research has explored what visual questions blind people would like to have answered~\cite{brady2013visual} and provided on-demand visual question answering support using both crowdsourcing~\cite{bigham2010vizwiz,kim2023exploring} and automated methods~\cite{gurari2018vizwiz}.
While VQA provides control over visual information gathering, it takes effort to ask individual questions. We investigate what types of visual questions BLV creators ask to create images during our formative study (similar to Brady et al.~\cite{brady2013visual}), then use VQA to extract visual information and summarize this information as image descriptions. Thus, we explore how VQA and image descriptions work together as interconnected rather than separate accessibility solutions.}

\subsection{Text-to-Image Generation Tools} 
In recent years, significant progress has been made in the field of generative image models, particularly text-to-image models. These models employ pre-trained vision-language models to encode text input into guiding vectors for image generation, allowing users to create images using text prompts. This advancement can be attributed to various factors, including innovations in deep learning architectures (e.g., Variational Autoencoders (VAEs)~\cite{kingma2013auto} and Generative Adversarial Networks (GANs)~\cite{goodfellow2020generative}), novel training paradigms like masked modeling for language and vision tasks~\cite{vaswani2017attention, devlin2018bert, dosovitskiy2020image, brown2020language}, and the availability of large-scale image-text datasets~\cite{schuhmann2022laion}.
With these advancements, recent diffusion-based models like DALL-E 2~\cite{ramesh2022hierarchical}, Stable Diffusion~\cite{rombach2021highresolution}, and Midjourney~\cite{midjourney} have successfully demonstrated the ability to synthesize high-quality images in versatile styles, including photorealism. This opens up potential practical applications for the content production industry~\cite{liu2022opal}. However, none of the image generation tools provide text descriptions of the output so they are not accessible to BLV creators. In this work, we chose to use MidJourney due to its popularity among designers and content creators for its high-quality results. MidJourney enables creators to generate 4 candidate images for a single text prompt via a text-based interface hosted on Discord. 
However, our approach is not limited to any particular model, as we focus on comparing and describing multiple generated results from a single prompt, helping creators select the ideal image from various candidates produced by image generation tools.

With the development of these models, recent works have conducted studies to understand the relationship between content creators and AI generative tools, introducing design guidelines for such systems~\cite{liu2022design, ko2023large}. These guidelines emphasize the need for more user controllability. Researchers have thus developed various tools to help designers better make use of generative AI, including assistance in exploring and writing better prompts~\cite{liu2022design, wang2023reprompt}, recommending potential illustrations for news articles~\cite{liu2022opal}, and supporting collaboration between writers and artists~\cite{ko2022we}. While these studies offer valuable insights into how designers interact with generative models, none have focused on creators with disabilities. Given the potential of text-to-image models for BLV creators, our work is the first to explore how to increase inclusivity in the expressiveness of image generation tools and make this emerging authoring approach more broadly accessible.

\section{Formative Study}
To understand the strategies and challenges of authoring and searching for images, we conducted a formative study with BLV creators. 
The formative study consisted of a semi-structured interview to investigate current strategies and challenges of obtaining images, and two image generation tasks to explore current strategies and challenges of using text-to-image generation.
  

\subsection{Method}
We recruited 8 BLV creators who create or use visual assets on a regular basis (P1-P8, Table~\ref{tab:form_participants}). Participants were recruited using mailing lists and compensated 50 USD for the 1.5-hour remote study conducted via Zoom\footnote{This study was approved by our institution's Institutional Review Board (IRB).}. 
Participants were totally blind (6 participants) or legally blind (2 participants) with light and color perception. All participants had previously produced or selected images for their work across several professions: teacher (English, Music), professor (Computer Science, Climate), software engineer, graduate student, and artist. 7 participants had prior knowledge of text-to-image generation models, none had previously used such tools.

We first conducted a semi-structured interview asking participants how they currently created or used visual assets, and what accessibility barriers they encountered with their current approaches. We then provided a short tutorial on text-to-image generation and shared Midjourney's guidelines for creating text prompts~\cite{midjourney_prompts} and example prompts from a Midjourney dataset~\cite{midjourney_dataset}. Participants then completed two image generation tasks (20 minutes per task): a \textit{guided task} in which participants generated a cover image for a news article~\cite{formative_article} given the article's title and full text, and a \textit{freeform task} in which participants generated their own image. 
To limit onboarding time, participant emailed us their prompt (text and/or image) instead of using Midjourney's Discord interface, then we shared the four generated candidate images back to the participants. We encouraged participants to ask questions about the four candidate images to select one or change the prompt. 
We recorded and transcribed the formative studies. To analyze the types of visual questions asked in the image generation task, two of the researchers labeled questions based on their goals and the types of information asked.\footnote{See Supplemental Material for the full list of prompts, images, and visual questions of the formative study}

\subsection{Findings}

\ipstart{Current Practice}
Participants reported that they currently use images for a variety of contexts including slides, website images, paintings for commission, cartoons, scientific diagrams, and music album covers (Table~\ref{tab:form_participants}). Five participants noted that they created images on their own using image creation software such as SVG editors, slides, photoshop, and ProCreate (P7, P1, P5, P6), code packages including Python and Latex (P4, P5), or by taking photos (P3). Among them, three participants asked sighted people to review them (P3, P4, P6), and two participants reviewed the images using accessibility tools (\textit{e.g.}, audioScreen, tactile graphs, ZoomText) (P7, P3). 
Five participants searched for images online (P7, P8, P2, P3, P5), and three participants recruited another person to create or search the images for them (P7, P4, P5). 

All participants who searched for images mentioned that they ask sighted people to describe the images for them in addition to reading any available alt text. P7 noted \textit{``Alt text has never been helpful. It's too short without important details.''}
P8 and P5 mentioned that while a few established websites (\textit{e.g.,} New York Times, NASA) have good alt text, Google Image Search returns options other than established websites and \textit{``it is hard to compare the results of the image search''} (P5). 
Participants also noted barriers to asking others to describe the image search results including finding available people to describe the images 
and avoiding false perceptions: \textit{``I only ask a handful of people because it might lead to some subconscious bias `that I’m not independent',  cause it’s a basic task''} (P7). 
\ipstart{Generating Prompts} All prompts written by participants specified the content they wanted to appear in the image (\textit{e.g.}, P6 used the prompt \textit{``A person pushing a grocery cart down a produce aisle.''}), and only two participants specified the style of the image (P1 and P7 specified \textit{``a photograph of...''}). Participants mentioned several challenges of creating prompts.
First, while prompt guidelines~\cite{midjourney_prompts} recommend users to specify multiple attributes in their prompt (\textit{e.g.}, style, lighting), participants reported that they were unfamiliar with visual attributes (\textit{``I’m trying not to leave much to system randomness, I want to detail more things. But I don't know a lot about different styles.''} --- P5) and others found it difficult to remember what to mention in the prompt: \textit{``I want the model to behave more like a wizard -- asking me a series of questions `What do you want to create?', `What style?' and so on. It is hard to create detailed prompts in one attempt} (P2). Participants also noticed that it is challenging to create a prompt that AI would be capable of generating: \textit{``If I pin down something really specific or narrow [in the prompt], AI seems to break down''} (P1). P5 mentioned that transparency could inform prompt iteration: \textit{``I want to know how the model works! [...] then I will know how to write a good prompt.''} Finally, while participants easily generated prompts during the free-form task motivated by their own creation goals, they mentioned it was challenging to know what content would effectively convey the article in the guided task: 
\textit{``I have no experience reading a news article with images, so it’s hard to think of one. What do these images usually contain?''} (P7).

\ipstart{Understanding Image Candidates\revised{with Visual Questions}}
After generating images, participants asked visual questions to understand and select the images. Participants asked a total of 89 questions (47 asked in the guided task, 42 in the freeform task).
The goals of the questions asked were to check whether the generated images followed the prompt (51), compare two or more images (34), request clarification of the answer provided by the interviewer (3), or understand a single image (1).   
The type of visual information asked by participants also varied. Participants asked about medium (5), settings (6), object presences (18), object types (11), position attributes (11), color/light/perspective (16), and others (22). 

Participants typically started by asking general questions, narrowing down to more specific questions as they ruled out images. For example, P4 progressively asked: \textit{``Can you describe the images?''}, \textit{``What are the differences between the four images?''}, \textit{``What are the differences between the [store] isles?''}, \textit{``Is the second image realistic?''}. Alternatively, participants started their questioning by directly checking if the image followed their prompts, such as in P5's first question: \textit{``Do we actually get the woman sitting at a desk?''} Finally, P1 and P2 started with questions about the style of the images: \textit{``Is it realistic or cartoony?''} (P1) and \textit{``Is the color calm or aggressive?''} (P2). 
Through asking questions, participants realized differences between their prompt and the generated images: \textit{``it seems like the model generator is filling in details according to the context, even if I didn’t specify some details. I didn’t specify the clothes but in all images, the women are wearing office clothes''} (P5). Participants then asked follow-up questions based on new details.
While the visual questions revealed the content and structure of what participants wanted to know about the images, participants reported that asking questions for each image was \textit{``very time-consuming and confusing''} (P4). 5 participants noted that they would prefer to receive descriptions before asking questions, and participants reported that remembering all of the answers was difficult, as P2 summarized: \textit{`` I wish there were more description provided in the first place. I don't know what to ask. Also, it's hard to remember all the answers for each image.''} 

\ipstart{Selecting an Image Candidate} While participants initially asked questions based on their prompt, they ultimately selected the final image considering both prompt-based descriptions and descriptions of extra details produced by the model. P7 suggested that information on whether the prompt is reflected in each image should be presented early so that he can decide whether to explore the image in detail or skip to the next candidate. P8 highlighted the importance of additional details: \textit{``The model has randomness. It showed items I didn’t ask for and didn’t show what I asked for in the prompt. I want much information to be surfaced so that I can make a decision. Whether that unexpected parts can be still used.''} 
We also observed that similarities between images guided participants in deciding whether to further explore the images or to refine the prompt. For instance, after P3 generated images using a prompt \textit{``A photo looking down on a kitchen table with a plate of pizza, a plate of fried chicken, and a bowl of ice cream on it.''}, he realized that all four images did not display drinks and iterated the prompt to explicitly mention ``fizzy drinks''. On the other hand, differences between the images ultimately informed the final selection, as participants cited unique backgrounds, objects, and mediums as reasons for selecting the image (\textit{e.g., P3 selected the final image because that was the only image that presented a dog putting his paw on the books.} ).
\ipstart{Uses of Image Generation} When participants generated their own images in the free-form task, participants created a variety of images ranging from logos, art, website decorative images, presentations, and music album cover.
All participants expressed excitement about using the text-to-image model as part of their image creation process in the future. 
Participants mentioned with image generation, they can create new types of images they had not created before. P6 mentioned \textit{``With SVG editor, I cannot make realistic images. But now I can!''} Also, participants mentioned that the quick creation will lead them to use images more often: \textit{``Because it's so quick, I will use it for communication. Similar to how sighted people draw on a whiteboard during a Zoom meeting, I can quickly generate an image because representing a concept visually is easier for sighted team members.''} (P8). P4 also compared the experience of image generation with image search ~\textit{``This simplifies things when I’m looking for things very niche, something that is hard to find online.''}
Finally, participants also mentioned the benefit of creating images alone. P7 said that because there is no need to ask a sighted person to help search images, it brings more autonomy and privacy. Participants also noted limitations and potential downsides of image generation including potential bias (P8, P4), copyright and training data concerns (P3, P4), wanting to use it only for inspiration (P1), and potential errors (P8). However, P8 expressed that he expected future models to produce fewer errors.


\subsection{Reflection}
Creators in our formative study currently employ resourceful strategies for creating or searching for images, but all creators expressed excitement to use image generation in their workflow. To improve access to image generation, our formative study reveals design opportunities (\textbf{D1}-\textbf{D5}) to make image generation accessible through technical or social support for: 
\begin{itemize}
    \item[\textbf{D1.}] Authoring prompts that specify content and style.
    \item[\textbf{D2.}] Understanding high-level image similarities and differences.
    \item[\textbf{D3.}] Assessing if images followed the prompt.
    \item[\textbf{D4.}] Accessing image details not specified by the prompt.
    \item[\textbf{D5.}] Organizing responses to visual questions.
\end{itemize}
These design opportunities address key user tasks in accessible text-to-image generation: generating the prompt (D1), understanding and selecting images (D2, D3, D4, D5), and revising the prompt for iteration (D4). Our work aims to help creators understand their image generation results through prompt-guided descriptions and comparisons (D2-D5). While providing high-quality descriptions may help creators improve their future prompts (D1), future work should explore how to actively support creators in authoring prompts.




\section{System}
We present \sysname{}, a system that supports accessible image generation via prompt-guided image descriptions and comparisons (Figure~\ref{fig:teaser}). 
To illustrate \sysname{}, we follow Vito, a professional blogger who uses a screen reader to author his articles. Vito recently wrote an article about the benefits of teaching children to cook, and he wants to add an image to the article to engage his sighted readers. He attempts to use image search to find a stock photo of \textit{``a young chef''} but notices that many of the images are missing detailed captions and alt text, or feature adult chefs instead of children. He decides to create an image using text-to-image generation with the prompt \textit{``a young chef is cooking dinner for his parents''}. The text-to-image generation model returns four candidates:
\begin{center}
  \centering
  \includegraphics[width=3.33in]{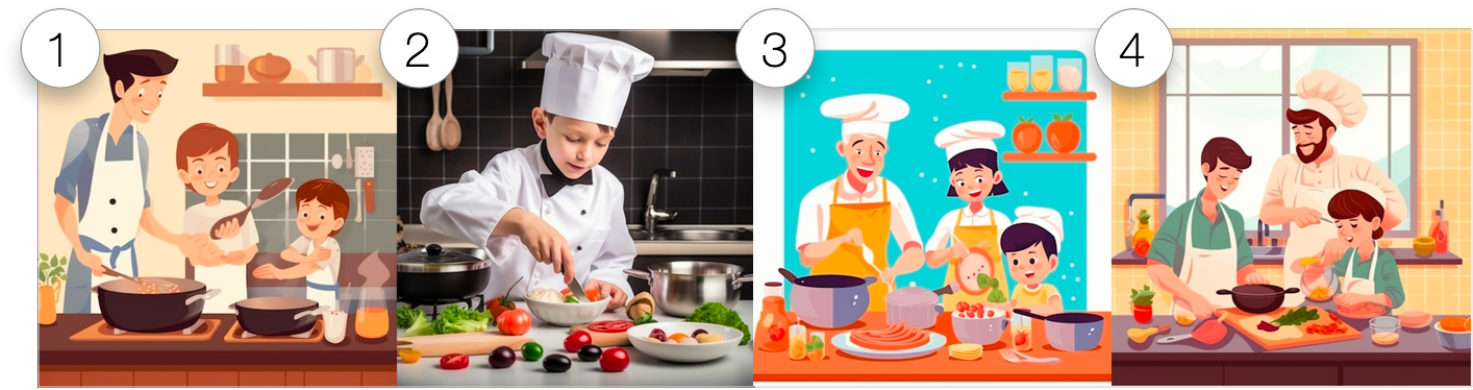}
  \label{fig:image_generation}
  \Description{
  Four images were generated using the prompt ``a young chef is cooking dinner for his parents.'' The first image is an illustration of a dad cooking with his two children. The second image is a photo of a young boy cooking alone in the kitchen. The third image is a vector art image that depicts parents and their sons cooking. The fourth image is an illustration of parents and their son who is a young man cooking in the kitchen with a wide window.
  }
\end{center}
\vspace{-10pt}
To decide whether to use one of these images or change his prompt, Vito enters his prompt and image results into \sysname{}.


\subsection{Prompt Verification}\label{sec:prompt_verification}
While the text-to-image model generates output images based on the prompt, the generated image often does not reflect the specifications in the prompt, especially if the prompt is long, complicated, or ambiguous~\cite{hu2023tifa}. 
To help users assess how well their generated images adhered to their prompt, \sysname{} provides \textbf{prompt verification}. 
To perform prompt verification, we first use GPT-4~\cite{openai2023gpt4} to generate visual questions that verify each part of the prompt. We input the text instruction \textit{``Generate visual questions that verify whether each part of the prompt is correct. Number the questions.''} followed by the user's prompt. GPT-4 outputs a series of questions:
\begin{center}
  \centering
  \vspace{1pt}
  \includegraphics[width=3.33in]{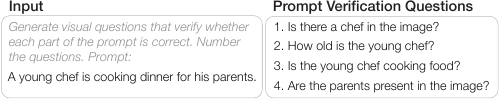}
  \label{fig:prompt_verification}
  \Description{A figure that illustrates an example of prompt verification questions. On the left is the input with the instruction ``Generate visual questions that verify whether each part of the prompt is correct. Number the questions.'' and the input prompt ``A young chef is cooking dinner for his parents.'' On the right is the output prompt verification questions: 1. Is there a chef in the image? 2. How old is the young chef? 3. Is the young chef cooking the food? 4. Are the parents present in the image?}
\end{center}
\vspace{-10pt}

\noindent We generate answers to the visual prompt verification questions for each of the four generated candidate images using the BLIP-2 model with the ViT-G Flan-T5-XXL setup~\cite{li2023blip}.
For each generated image and prompt verification question, we instruct the BLIP-2 model with the starting sequence \textit{``Answer the given question. Don’t imagine any contents that are not in the image.''} to reduce hallucinations with non-existent information:
\begin{center}
  \centering
  \vspace{2pt}
  \includegraphics[width=3.33in]{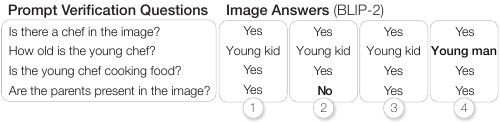}
  \vspace{-12pt}
  \label{fig:prompt_verification_per_image}
  \Description{A figure that illustrates an example of prompt verification questions and their answers for each of the four images. The first question is ``Is there a chef in the image?'' and the answer is all yes for four images. The second question is ``How old is the young chef?'' and the first three images answer ``Young kid'' while the last image says ``Young man''. The third question is ``Is the young chef cooking food?'' and the answers are all yes for the four images. The final question is ``Are the parents present in the image?'' and the answer is all yes except for the second image.}
\end{center}
To help users quickly find which images do or do not adhere to the prompt, we use GPT-4 to summarize the responses to each question using the following prompt: \textit{``Below are the answers of four similar images to one visual question. Write one sentence summary that captures the similarities and differences of these results. The summary should fit within 250 character limit''}. When using GPT-4's chat completion API, we set the role of the system as \textit{``You are a helpful assistant that is describing images for blind and low vision individuals.''}. The temperature value was set to 0.8. The summaries either indicate that all images have the same answer (\textit{e.g.}, ``All images have a chef in the image''), or they alert users to differences:
\begin{center}
\vspace{+1pt}
  \centering
  \includegraphics[width=3.33in]{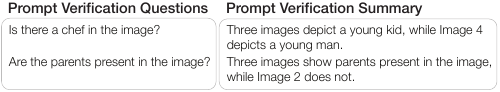}
  \vspace{-14pt}
  \label{fig:prompt_verification_summary}
  \Description{This figure illustrates the example of the summary descriptions for the prompt verification questions. For the question ``Is there a chef in the image?'', the summary description is ``Three images depict a young kid, while Image 4 depicts a young man.'' For the second question ``Are the parents present in the image?'', the answer summary is ``Three images show parents present in the image, while image 2 does not.''}
\end{center}

\begin{figure}
  \centering
  \includegraphics[width=3in]{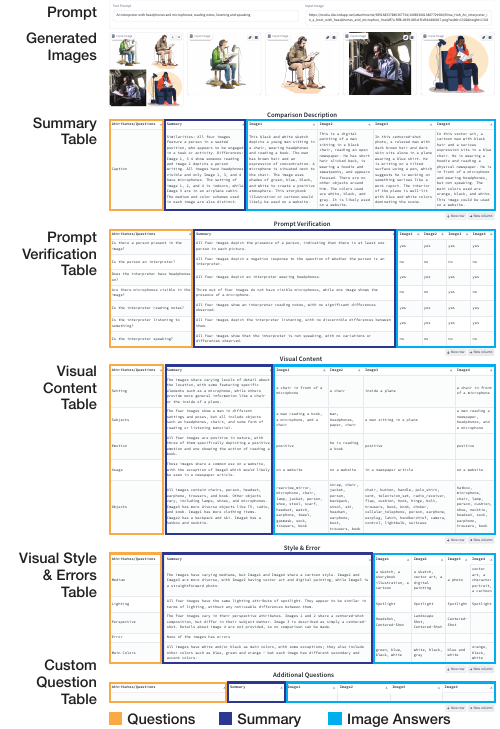}
  \caption{The \sysname{} interface consists of screen reader accessible tables that enable users to flexibly gain more information about the content of interest.}
  \label{interface}
  \Description{The screenshot of the actual GenAssist Interface. On top are the prompt and images, followed by the summary table, prompt verification table, content table, style and errors table, and custom question tables.}
\end{figure}

To enable screen reader users to easily access the answers to each question, we present the prompt verification results as a table including the prompt verification questions (rows, with the question in column \#1), prompt verification summaries (column \#2), and per-image prompt verification answers (columns \#3-6) (Figure~\ref{interface}). 

Using our prompt verification table, Vito reads the answers summaries to check if the images follow his prompt. He notices that the 4th image contains an older chef, so it does not apply to his article about teaching children how to cook. While Vito also realizes the 2nd image does not feature the chef's parents, he keeps the image in consideration as it may still apply to his article.



\subsection{Visual Content \& Style Extraction}
Generated image candidates often feature similarities or differences that are not present in the original prompt. For example, Vito's prompt \textit{``A young chef is cooking dinner for his parents''} does not specify the style such that the resulting images include three illustrations and one photo. 
To enable access to image content and style details that were not specified in the prompt, we extract the \textbf{visual content} and \textbf{visual style} of the generated image candidates. To surface content and style similarities and differences that are important for improving image generation prompts, we used text-to-image prompt guidelines~\cite{midjourney_prompts, midjourney_styles, dalle_promptbook} to inform our approach. 

We first created a list of visual questions about the image based on existing prompt guidelines, i.e. \textit{prompt guideline questions}. The prompt guideline questions consist of questions about the content of the image (subjects, setting, objects), the purpose of the image (emotion, likely use), the style of the image (medium, lighting, perspective, color), and an additional question about errors in the image to surface distortions in the generated images such as blurring or unnatural human body features (Table~\ref{tab:question-table}). 

To answer our prompt guideline questions for each image, we answered 5 questions (setting, subjects, emotion, likely use, colors) using Visual Question Answering with BLIP-2, similar to our prompt verification approach: 
\vspace{-2pt}
\begin{center}
    \centering
    \includegraphics[width=3.33in]{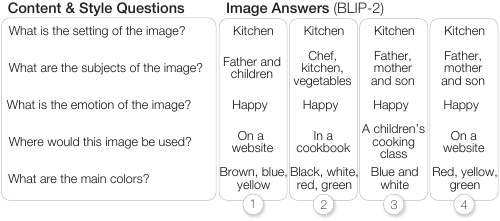}
    \vspace{-10pt}
    \Description{This image illustrates an example of content and style questions answered by BLIP-2. The answers are provided for the same tutorial images. For the question ``What is the setting of the image?'', the answers are all kitchen. For the question ``What are the subjects of the image?'', the answers are father and children for the first image, chef, kitchen, and vegetables for the second image, and father, mother, and son for the third and fourth images. For the question ``What is the emotion of the image?'', all answers are happy. For the question about the usage ``Where would this image be used?'', the answers are ``on a website'', ``in a cookbook'', ``a children's cooking class'', and ``on a website''. Finally, for the question ``What are the main colors?'', the first image answers ``Brown, blue, yellow'', the second image answers ``Black, white, red, green'', the third image answers ``blue and white'', and the final image answers ``Red, yellow, green''.}
\end{center}
\noindent For our objects question, we used Detic~\cite{zhou2022detecting}, a state-of-the-art object detection model, with an open detection vocabulary and a confidence threshold of 0.3 to enable users to access all objects:
\begin{center}
    \centering
    \includegraphics[width=3.33in]{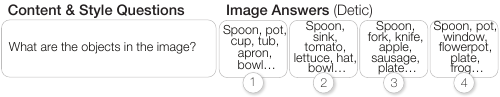}
    \vspace{-15pt}
\Description{This figure shows an example of the object detection results using Detic per image. The first image depicts a spoon, pot, cup, tub, apron, bowl, etc. The second image depicts a spoon, sink, tomato, lettuce, hat, bowl, etc. The third image shows a spoon, fork, knife, apple, sausage, plate, etc. The fourth image shows a spoon, pot, window, flowerpot, plate, frog, etc.}
\end{center}

\noindent For the remaining questions covering medium, lighting, perspective, and errors, we answer the question for each image candidate by using CLIP~\cite{radford2021learning} to determine the similarity between the image and a limited set of answer choices (similar to CLIP interrogator~\cite{interrogator}). To provide answers that could inform future prompts, we curated our answer choices for medium, lighting, and perspective from Midjourney's list of styles~\cite{midjourney_styles} and DALL-E's prompt book~\cite{dalle_promptbook}. To address common image generation errors, we retrieved the answer choices for our errors question from prior work~\cite{reddy2021dall, stable_diffusion_negativeprompts}. We include the full list of answer choices in the Supplementary Material. For each question, \sysname{} presents the top three answer choices with a similarity score between the answer choice embedding and the image embedding above a threshold of 0.18:
\begin{center}
    \centering
    \includegraphics[width=3.33in]{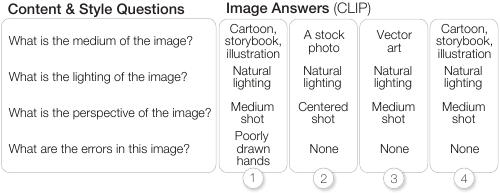}
    \vspace{-18pt}
    \Description{This figure shows the answers related to the content and style of the images that were retrieved using the CLIP model. First for the medium of the image, image 1 answers cartoon, storybook, and illustration, image 2 answers a stock photo, image 3 answers vector art, and image 4 answers cartoon, storybook, and illustration. For the lighting of the image, all four images answer natural lighting. For the perspective of the images, images 1, 3, and 4 answer medium shots while the second image answers centered shots. For the errors in the image, only image 1 answers poorly drawn hands but not the other three images.}
\end{center}
To inform creators about unfamiliar visual style types, \sysname{} provides the definition and the usage for each answer choice for visual style questions (Medium, Lighting, Perspective) by generating the description with GPT-4 and the prompt \textit{``Describe the definition and the usage of the following [QUESTION NAME] in one sentence: [STYLE NAME]''.} Similar to the prompt verification table, we present the prompt guideline results in a table format including the prompt guideline questions (rows, with the question in column \#1), prompt guideline summaries (column \#2), and per-image prompt guideline answers (columns \#3-6). We further split the prompt guideline results into two tables to improve ease of navigation: the \textit{visual content table} includes answers to the content and purpose questions, and the \textit{visual style table} includes answers to the style and errors questions.
Finally, users can ask their own questions at the bottom of either table and \sysname{} adds a row to the table by generating the answer for each image using BLIP-2, and the summary of answers using GPT-4.
\begin{table}
\resizebox{3.33in}{!}{%
\begin{tabular}{@{}llll@{}}
\toprule
Category & Name       & Question                                     & Model  \\ \midrule
Content  & Setting     & What is the setting of the image?            & BLIP-2 \\
         & Subjects    & What are the subjects of the image?          & BLIP-2 \\
         & Objects     & What are the objects in this image?          & Detic \\  
          & Emotion     & What is the emotion of the image?            & BLIP-2 \\
         & Usage       & Where would this image likely be used?       & BLIP-2 \\ \midrule
Style    & Medium      & What is the medium of the image?            & CLIP   \\
\& Errors & Lighting    & What is the lighting in this image?          & CLIP   \\
         & Perspective & What is the perspective of this image?       & CLIP   \\
         & Colors      & What are the main colors used in this image? & BLIP-2 \\ 
    & Errors      & What are the errors in this image?           & CLIP   \\ \bottomrule
\end{tabular}%
}
\caption{Our \textit{prompt guideline questions} including the question category, question name, and question, along with the model we used to answer the question (BLIP-2~\cite{li2023blip}, CLIP~\cite{radford2021learning}, or Detic~\cite{zhou2022detecting}).}
\label{tab:question-table}
\Description{The table illustrates the type of questions and which models are used to answer the questions. In the Content category, setting, subjects, emotion, and usage information is retrieved using BLIP-2, and object information is retrieved using Detic. In the Style and Errors category, medium, lighting, perspective, and errors information is retrieved using CLIP, and the color information is retrieved using BLIP-2.}
\end{table}
Using the visual content table, Vito notices from the objects summary that Image 1 has more food items than Images 2-4. As the purpose of the article is partially to introduce children to more ingredients, he decides to remove Image 1 from consideration. Using the visual style table, Vito realizes that Image 2 is a photo, while the other images are illustrations. As Vito was initially searching for a photo, he notes he may want to further refine his prompt to get more photo results. Vito also wants to check if the images will match his blog which is primarily black and white, so he adds a question about the background color:
\begin{center}
  \centering
  \includegraphics[width=3.33in]{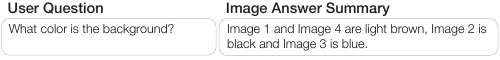}
  \label{fig:new_question_answer}
  \Description{This image shows the summary description of the additional question. When asked ``What color is the background?'', the summary description is ``Image 1 and Image 4 are light brown, Image 2 is black and Image 3 is blue.''}
\end{center}
\vspace{-15pt}
As Image 2 fits his article and includes a black background, he ranks Image 2 as his current top choice.


\subsection{Description Summarization}

To enable users to quickly assess their image results, we summarize the results from our pipeline to create a per-image description for each image and a summary of image similarities and differences.

To generate \textbf{per-image descriptions}, we first obtain the BLIP-2 caption for each image that provides a concise overview of the image content (\textit{e.g.,} \textit{``A family preparing food in the kitchen with a window.''}). Then, we obtain additional detail about the image by generating questions about the caption with GPT-4 with the prompt: \textit{``Given the caption, generate 10 visual questions that are likely to be asked by blind and low vision individuals''.} Unlike the other questions in our pipeline that are common across all images, this step enables the \sysname{} to ask image-specific questions to add detail (\textit{e.g.}, \textit{``What is the view outside the window?''} is only asked for Image 4). We generate the answers to these questions using BLIP-2. 

\begin{figure}[t]
  \centering
  \includegraphics[width=3.33in]{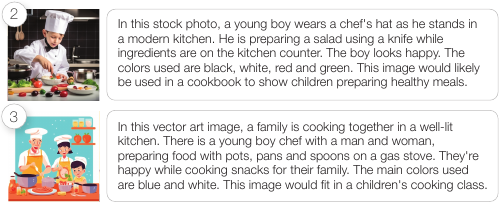}
  \vspace{-10pt}
  \caption{\sysname's per-image descriptions.}
  \label{fig:rich}
  \Description{This figure illustrates the per-image descriptions of the second and third images from the tutorial image sets. The first description is: In this stock photo, a young boy wears a chef's hat as he stands in a modern kitchen. He is preparing a salad using a knife while the ingredients are on the kitchen counter. The boy looks happy. The colors used are black, red, and green. This image would likely be used in a cookbook to show children preparing healthy meals. The second description is: In this vector art image, a family is cooking together in a well-lit kitchen. There is a young boy chef with a man and woman, preparing food with pots, pans, and spoons on a gas stove. They're happy while cooking snacks for their family. The main colors used are blue and white. The image would fit in a children's cooking class.} 
\end{figure}

We create individual image descriptions by first aggregating all information acquired in our pipeline for each image including the prompt verification, prompt guideline, and caption-detail question-answer pairs for each image. Then, we guide GPT-4 with the aggregated visual information and the prompt \textit{``Below is the information of an image. Write a description of this image for the blind and low vision audience. Describe the medium first. Your response should fit within 250 character limit. Do not add additional information that was not provided. Do not describe parts that are not clear or cannot be determined from the given information.''} GPT-4 generates rich descriptions for each image (Figure~\ref{fig:rich}). 

\begin{figure}
  \centering
  \includegraphics[width=3.33in]{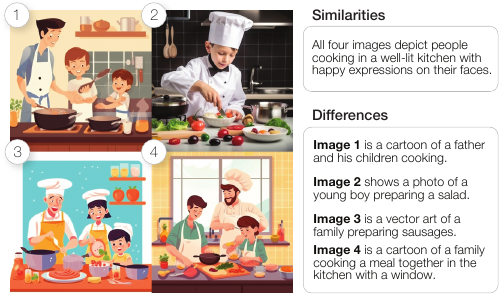}
  \caption{\sysname's image comparison descriptions.}
  \label{fig:comparison}
  \Description{This figure shows the example of image comparison descriptions: the similarities and differences between the four images. Similarities: All four images depict people cooking in a well-lit kitchen with happy expressions on their faces. Differences: Image 1 is a cartoon of a father and his children cooking. Image 2 shows a photo of a young boy preparing a salad. Image 3 is a vector art of a family preparing sausages. Image 4 is a cartoon of a family cooking a meal together in the kitchen with a window.}
  \vspace{-10pt}
\end{figure}

\noindent To generate the \textbf{comparison description}, we simply provide all the information extracted from our pipeline to GPT-4 with the prompt \textit{``Below is the information for four images. Write one paragraph about the similarities between the four images and one paragraph about the differences between the four images. The summary should be concise.''}. GPT-4 briefly summarizes the image similarities and differences (Figure~\ref{fig:comparison}).
\noindent To help users quickly assess whether to revise their prompt or continue exploring, we present the \textbf{comparison description} and \textbf{per-image description} at the top of the page before the prompt verification and prompt guidelines tables.

With the per-image description, Vito can quickly recall the content of Image 2 before making his final selection. With the comparison descriptions, Vito can quickly notice that Image 2 was the only image that contained a photo, then updated his prompt to get additional photos rather than illustrations.

\subsection{Implementation}
We implemented \sysname{} using Gradio~\cite{abid2019gradio}, an open-source Python library for the front-end web interface. The interface was deployed through Hugging face~\footnote{https://huggingface.co/spaces} space with an NVIDIA A100 GPU (large, 40GB GPU Memory). Uses' interaction logs were saved in the Firebase database. We followed the guidelines of W3C~\cite{wai} and tested the compatibility of the \sysname{} with all three major screen readers: \textit{NVDA}, \textit{JAWS}, and \textit{VoiceOver}. \sysname's tables follow the recommendations of W3C tables with two headers~\footnote{https://www.w3.org/WAI/tutorials/tables/two-headers/}.

\section{Pipeline Evaluation}
We measured the \textit{coverage} of the descriptions generated by \sysname{} and the \textit{accuracy} of the information presented in \sysname{}'s tables. We compare the coverage of \sysname{}-generated caption with the human-generated caption and the caption generated by a state-of-the-art image captioning model BLIP-2~\cite{li2023blip}.

\begin{figure*}[htbp!]
  \centering
  \includegraphics[width=0.95\textwidth]{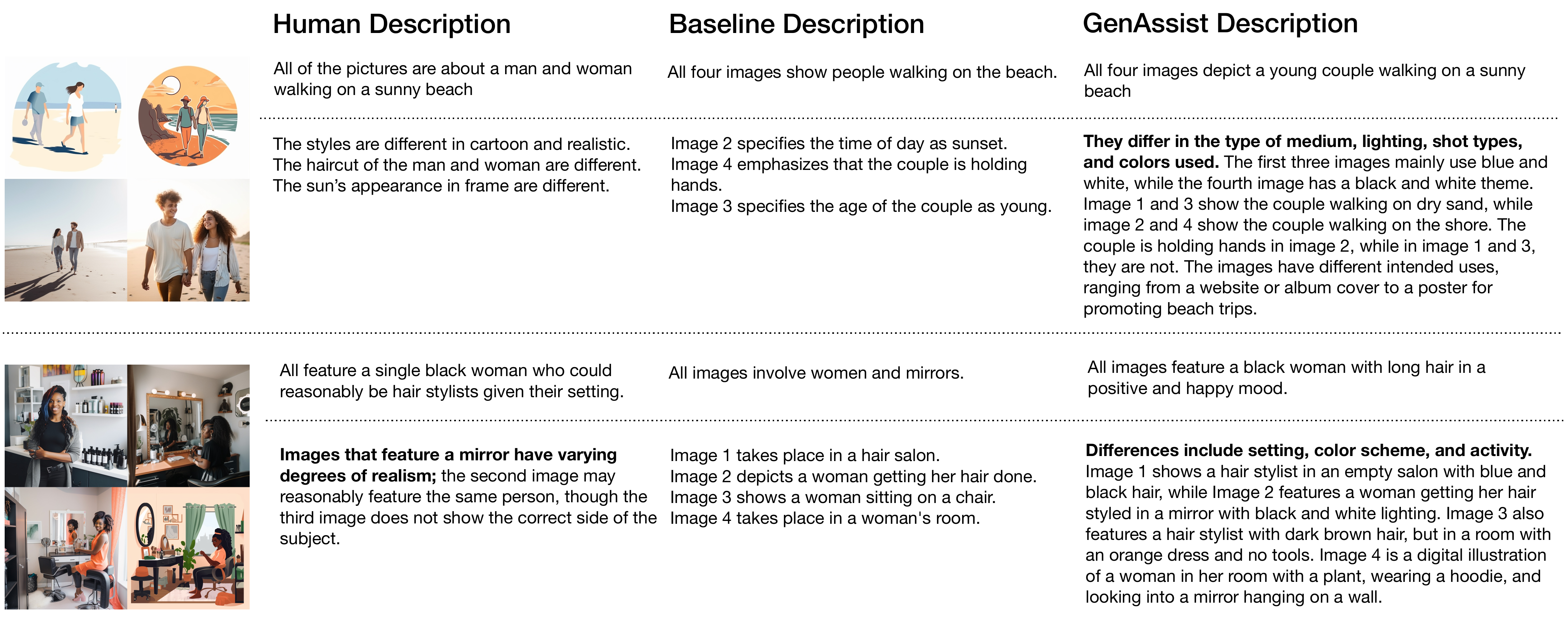}
  \caption{Two image sets and the descriptions of the similarities and differences used in the pipeline coverage evaluation (each image set described by a different human describer). \sysname{} captured more information in the similarities and differences caption than the human describers.
  }
  \label{fig:pipeline_coverage}
  \Description{The figure compares the description generated by humans, the baseline, and the system with two image sets. First set by human: All of the pictures are about a man and woman walking on a sunny beach. The styles are different in cartoon and realistic. The haircuts of the man and woman are different. The sun’s appearance in the frame is different. First set by baseline: All four images show people walking on the beach. Image 2 specifies the time of day as sunset. Image 4 emphasizes that the couple is holding hands. Image 3 specifies the age of the couple as young. First set by system: All four images depict a young couple walking on a sunny beach. They differ in the type of medium, lighting, shot types, and colors used. The first three images mainly use blue and white, while the fourth image has a black-and-white theme. Images 1 and 3 show the couple walking on dry sand, while images 2 and 4 show the couple walking on the shore. The couple is holding hands in image 2, while in images 1 and 3, they are not. The images have different intended uses, ranging from a website or album cover to a poster for promoting beach trips. Second set by human: All feature a single black woman who could reasonably be hair stylist given their setting. Images that feature a mirror have varying degrees of realism; the second image may reasonably feature the same person, though the third image does not show the correct side of the subject. Second set by baseline: All images involve women and mirrors. Image 1 takes place in a hair salon. Image 2 depicts a woman getting her hair done. Image 3 shows a woman sitting on a chair. Image 4 takes place in a woman's room. Second set by system: All images feature a black woman with long hair in a positive and happy mood. Differences include setting, color scheme, and activity. Image 1 shows a hair stylist in an empty salon with blue and black hair, while Image 2 features a woman getting her hair styled in a mirror with black and white lighting. Image 3 also features a hair stylist with dark brown hair but in a room with an orange dress and no tools. Image 4 is a digital illustration of a woman in her room with a plant, wearing a hoodie, and looking into a mirror hanging on a wall.}
\end{figure*}

\subsection{Method}
We selected 20 image sets (20 prompts x 4 generated images for each prompt = 80 total images) from Midjourney's community feed spanning different prompt lengths, content types, and styles. We recruited two people with experience describing images to provide descriptions for 10 randomly selected image sets each. For each image set, the describers provided descriptions of each individual image, and the similarities and differences between the images.
We provided describers with prompt guidelines~\cite{midjourney_prompts}, image description guidelines~\cite{image_desc_guideline}, an example set of descriptions created by \sysname{}, and the prompt for each image set to inform their descriptions. Both describers spent 3.5 hours to create descriptions for the 10 sets of images --- or around 21 minutes per image set. 

We compared the coverage of \sysname{}-generated descriptions to those generated by a baseline captioning tool (BLIP-2) and human describers. For comparison, we annotated the similarities and differences descriptions for all 20 sets of images and annotated the individual descriptions for 10 sets of images. We chose the 10 sets with the longest human descriptions to compare \sysname{} with the highest quality descriptions.
Because BLIP-2 cannot take multiple images as input to extract similarities and differences, we generated captions of the 4 images using BLIP-2, then prompted GPT-4 with the same prompt we used in our system to generate summary descriptions.
We tallied whether the descriptions contained details about the image in each of our set of pre-defined visual information categories (Table~\ref{tab:question-table}). 
We counted only the correct information in the descriptions.
One of the researchers annotated the descriptions and the other researcher reviewed the annotations. 
To compute the accuracy of the detailed visual information in \sysname{}, one of the researchers examined the 20 sets of images with the three tables generated by the \sysname{} (prompt verification table, visual content table, and visual style table) and counted the number of correct and incorrect answers in each table. 

\begin{table}
\small\sffamily\def\arraystretch{0.75}\setlength{\tabcolsep}{0.5em}
\centering{%
\begin{tabular}{@{}llll@{}}
\toprule
Category & Sub-category & Correct (\%)                                     & Correct (\#)  \\ \midrule
Prompt verification  &  & 92.82       & 418 \\ \midrule
Content  & Setting     & 97.53            & 81 \\
         & Subjects    & 98.60          & 143 \\
         & Objects    & 82.86          & 1243 \\ 
  & Emotion     & 87.5            & 80 \\
         & Usage       & 97.50       & 80 \\ \midrule
Style    & Medium      & 82.76             & 174   \\
         & Lighting    & 94.33          & 141   \\
         & Perspective & 71.83       & 142   \\
         & Colors      & 99.1  & 221 \\ \midrule
Errors   &       & 60.00           & 5   \\ \bottomrule
\end{tabular}%
\caption{\revised{We report the accuracy (percentage and number of correctly predicted information) of the pipeline results (Prompt verification, Content, Style, and Errors) with 20 sets of images.}}
\vspace{-20pt}
    \label{tab:pipeline_accuracy}
    \Description{This table shows the accuracy values of the pipeline results of the GenAssist with 20 sets of images.}
}
\end{table}

\begin{table*}[t]
\small\sffamily\def\arraystretch{0.8}\setlength{\tabcolsep}{0.4em}
\centering
\begin{tabular}{lrrrrrrrrrr|rrr}
\toprule
\multicolumn{2}{l}{\multirow{2}{*}{(Correct Only)}} & \multicolumn{3}{c}{Total Content (\#)} & \multicolumn{3}{c}{Total Style (\#)} & \multicolumn{3}{c}{Total Error (\#)}  & \multicolumn{3}{|c}{Total (\#)} \\
\multicolumn{2}{c}{} & \multicolumn{1}{l}{Human} & \multicolumn{1}{l}{Baseline} & \multicolumn{1}{l}{\sysname{}} & \multicolumn{1}{l}{Human} & \multicolumn{1}{l}{Baseline} & \multicolumn{1}{l}{\sysname{}} & \multicolumn{1}{l}{Human} & \multicolumn{1}{l}{Baseline} & \multicolumn{1}{l}{\sysname{}} & \multicolumn{1}{|l}{Human} & \multicolumn{1}{l}{Baseline} & \multicolumn{1}{l}{\sysname{}} \\ \midrule
{\multirow{2}{*}{Similarities}} & \textbf{$\mu$} & 1.5 & 1.65 & \textbf{2.45} & 0.70 & 0.00 & \textbf{0.80} & \textbf{0.10} & 0.00 & 0.00 & 2.35 & 1.65 & \textbf{3.25} \\
\multicolumn{1}{r}{} & \textbf{$\sigma$} & 0.61 & 0.59 & 1.10 & 0.80 & 0.00 & 0.83 & 0.31 & 0.00 & 0.00  & 0.83 & 0.85 & 1.29\\ \midrule
\multirow{2}{*}{Differences} & \textbf{$\mu$} & 1.50 & 1.95 & \textbf{2.35} & 0.65 & 0.35 & \textbf{2.20} & \textbf{0.05} & 0.00 & 0.00 & 2.25 & 2.30 & \textbf{4.55}\\
 & \textbf{$\sigma$} & 0.69 & 0.39 & 0.49 & 0.75 & 0.49 & 1.01 & 0.22 & 0.00 & 0.00 & 0.84 & 0.93 & 1.26\\\midrule
Per-Image & \textbf{$\mu$} & \textbf{1.71} & 0.69 & \textbf{1.71} & \textbf{0.71} & 0.04 & 0.68 & \textbf{0.05} & 0.00 & 0.01 & \textbf{2.47} & 0.73 & 2.41\\
Descriptions & \textbf{$\sigma$} & 0.39 & 0.10 & 0.26 & 0.22 & 0.07 & 0.30 & 0.05 & 0.00 & 0.03 & 0.74 & 0.33 & 0.75 \\
 \bottomrule
\end{tabular}
\caption{We compared the coverage of \sysname{}-generated descriptions to those generated by a baseline captioning tool and human describers. \sysname{} captured more similarities and differences than the human describers. }
    \label{tab:pipeline_coverage}
    \Description{Table's rows represent similarities, differences, and per-image descriptions, and the columns represent the number of content, style, error, and total visual information captured by human, baseline, and GenAssist. For similarities and differences, GenAssist's coverage is higher than the other two conditions except for error detection. For per-image description, GenAssist's and human-generated description's coverage are comparable.}
    \vspace{-5pt}
\end{table*}

\subsection{Results}
\subsubsection{Coverage}~\label{sec:pipeline_coverage}
\revised{We summarize our coverage evaluation results in \autoref{tab:pipeline_coverage}. Overall, \sysname{}'s comparison descriptions covered more similarities and differences than the human describers'. In the coverage of differences, \sysname{} spotted more than twice the number of total differences than the human describers (4.55 vs. 2.25). The coverage of \sysname{}'s individual image descriptions was comparable to that of human describers. 
When compared to human-generated description, \sysname{} captured more information about the content and styles but revealed fewer image generation errors. For instance, one human describer specified in the comparison description \textit{``...All of the images have some AI generation error with fingers or clothing. 
''}.
While \sysname{} and the baseline used the same GPT-4 prompt to extract the similarities and differences, the baseline's comparison description did not capture many differences.}

\subsubsection{Accuracy}
\autoref{tab:pipeline_accuracy} summarizes the results of the accuracy evaluation. Prompt verification, content, and style categories all achieved over 90\% accuracy except for medium, perspective and emotion.
In the 80 images in the dataset, \sysname{} only detected five images as having errors, and detected the correct error types in three of them. 
The most common errors made in our pipeline were from perspective, medium, and error categories which are all extracted using the CLIP score. For perspective and medium, the majority of the errors were due to CLIP matching images to common style expressions (\textit{e.g.,} natural lighting, centered-shot) which likely reflects prevalence of these expressions in the training data. In the incorrect output of errors, \sysname{} detected cartoon or sketch images as \textit{`poorly drawn faces'} errors. 
One reason for the relatively low accuracy of object detection results is that we empirically set the output threshold of \sysname{}'s object detection (Detic) as 0.3 to present diverse objects to users in addition to information about the main subject extracted by BLIP-2 in our pipeline. 

\section{User Evaluation}
We conducted a user study with 12 BLV visual content creators to compare \sysname{} with a baseline interface.

\subsection{Method}
In a within-subjects study, participants used \textit{\sysname{}} and a \textit{baseline} interface to interpret image generation results (\textit{interpretation task}) and to generate images (\textit{generation task}). 

\ipstart{Participants} We recruited BLV creators who create or use visual assets on a regular basis using mailing lists (P7-P18, Table~\ref{tab:form_participants}). Participants described their vision as totally or legally blind and they were students, consultants, software engineers, video creators, and artists. P7 and P8 participated in the formative study.

\ipstart{Baseline} The baseline interface included for each image: the image caption from BLIP-2~\cite{li2023blip}, a list of objects from Detic~\cite{zhou2022detecting}, and the ability to interactively ask visual questions powered by BLIP-2~\cite{li2023blip}. We designed the baseline to encompass commonly used captioning and object detection tools available in commercial devices and applications (\textit{e.g.}, SeeingAI~\cite{seeingAI}). As such captions tend to be concise, we added visual question answering via BLIP-2~\cite{li2023blip} to let participants gain additional information on-demand.

\begin{figure}
  \centering
  \includegraphics[width=3.33in]{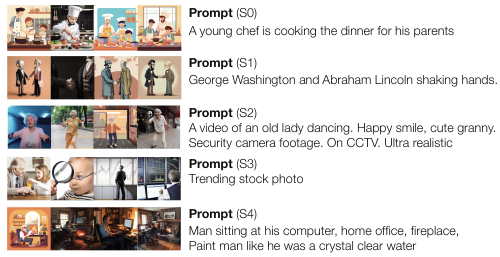}
  \caption{We selected two sets of images from Midjourney's community feed generated with a short prompt without detailed descriptions of objects or styles (S1, S3) and two sets with a long prompt with detailed descriptions of objects or styles (S2, S4). We selected long and short prompts to explore how users compared images when they are similar (long prompts) vs. dissimilar (short prompts).}
  \label{fig:materials}
  \Description{This figure shows the image sets used in our evaluation study. S0 is the set of images generated with the prompt ``A young chef is cooking the dinner for  his parents.'' S1 is the images generated with the prompt ``George Washington and Abraham Lincoln shaking hands.'' and S2 is the images generated with the prompt ``A video of an old lady dancing. Happy smile, cute granny. Security camera footage. On CCTV. Ultra realistic.'' S3 is generated with the prompt ``Trending stock photo'' and S4 is generated with the prompt ``Man sitting at his computer, home office, fireplace. Paint man like he was a crystal clear water.''}
  \vspace{-10pt}
\end{figure}

\ipstart{Procedure} 
We first asked participants demographic and background questions about how they use images in their work. We then gave a 15-minute tutorial on both the \sysname{} interface and the baseline interface using S0 (\autoref{fig:materials}). Participants then completed two tasks: the interpretation task and the generation task.

In the interpretation task, participants used both interfaces to evaluate pre-generated images (Figure~\ref{fig:materials}). For each set of images, we provided participants with an example scenario (\textit{e.g.,} Select an image for a blog post titled \textit{`My grandma still dances!'}). Using \sysname{} or the baseline interface, participants were asked to identify the similarities and differences in the image candidates and choose a final image. For each interface, users were given one short prompt image set (S1 or S3) and one long prompt image set (S2 or S4). The order of the interfaces and image sets were counterbalanced and randomly assigned to participants. After each interface, we conducted a post-stimulus survey that included the following ratings: Mental Demand, Performance, Effort, Frustration, and Usefulness of the caption in understanding differences between images. All ratings were on a 7-point Likert scale.

In the generation task, we provided participants with the title and first 5 paragraphs of two articles, then asked participants to create a relevant image for the article by coming up with their own prompts. We selected the two articles from the New York Times: \textit{`Why Multitasking is Bad for You}' and \textit{`My Kids Want Plastic Toys. I Want to Go Green.'}~\cite{multitasking, plastic}. The order of the interfaces and articles was counterbalanced and randomly assigned to participants. After each interface, we asked the participants to choose one image from the generated images and explain their reasoning. We also conducted a post-stimulus survey that included the following ratings: Mental Demand, Performance, Effort, Frustration, Usefulness of the caption, Satisfaction with the final image, and Confidence in posting the final image. All ratings were on a 7-point Likert scale. At the end of the study, we conducted a semi-structured interview to understand participants' strategies using ~\sysname{} and the pros and cons of both ~\sysname{} and the baseline.

The study was 1.5 hours long, conducted in a 1:1 session via Zoom, and approved by our institution's IRB. We compensated participants 50 USD for their time.

\ipstart{Analysis}
We recorded the study video, user-generated prompts and images, and the survey responses.
We transcribed the exit interviews and participants' spontaneous comments during the tasks and grouped the transcript according to (1) strategies of using \sysname{} and (2) perceived benefits and limitations of our system.

\subsection{Results}
\begin{figure*}[t]
  \centering
  \includegraphics[width=0.95\textwidth]{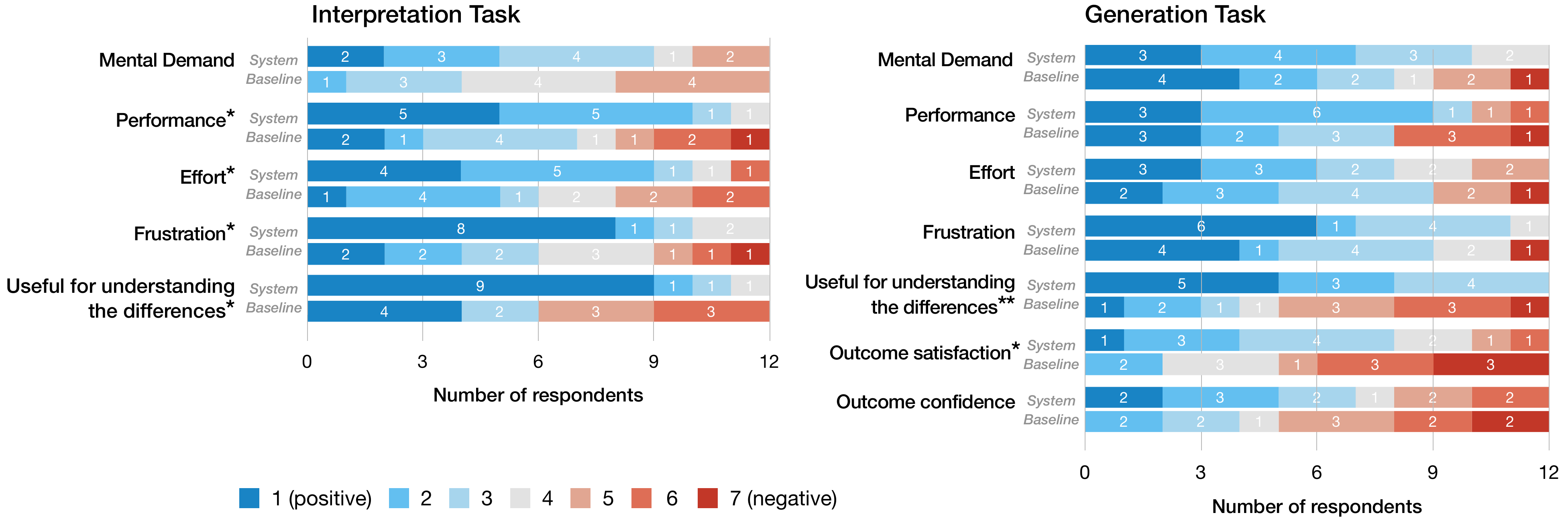}
  \caption{Distribution of the rating scores for \sysname{} and the baseline interface (1 = positive, 7 = negative) in the two tasks. Note that a lower value indicates positive feedback and vice versa. 
  {The asterisks indicate the statistical significance as a result of Wilcoxon text} (~\textit{p} < 0.05 is marked with * and ~\textit{p} < 0.01 is marked with **). In the interpretation task, \sysname{} significantly outperformed the baseline interface in performance, effort, frustration, and usefulness for understanding the differences between images. In the generation task, \sysname{} was significantly lower in being useful for understanding the differences and in outcome satisfaction.}\label{fig:likert_scale}
  \Description{Stacked bar char that visualizes the distribution of the rating scores for GenAssist and the baseline interface. Performance, effort, frustration, and usefulness for understanding the differences are significant in the interpretation task. Outcome satisfaction and the usefulness for understanding the differences are significant in the generation task.}
\end{figure*}

Overall, all participants stated they would like to use ~\sysname{} rather than the baseline interface to create images in the future. Participants expressed that \sysname{} would be immediately useful in their workflows: \textit{``This is usable out of the box!'' [...] ``I need access to this technology''} (P14), \textit{``I’d even pay for this! I really need this''} (P15).
In particular, participants rated \sysname{} to be significantly more useful for understanding the differences between images in both tasks (interpretation: $\mu$=1.50, $\sigma$=1.00 vs. $\mu$=3.58, $\sigma$=4.00; $Z$=-2.31; $p$<0.05; generation: $\mu$=1.92, $\sigma$=2.00 vs. $\mu$=4.33, $\sigma$=5.00; $Z$=-2.77; $p$<0.01) (Figure~\ref{fig:likert_scale}). 
For the interpretation task, participants reported significantly better performance ($\mu$=1.83, $\sigma$=2.00 vs. $\mu$=3.67, $\sigma$=3.00; $Z$=-2.47; $p$<0.05), significantly less frustration ($\mu$=1.75, $\sigma$=1.00 vs. $\mu$=3.50, $\sigma$=3.50; $Z$=2.46; $p$<0.05), and effort ($\mu$=2.25, $\sigma$=2.00 vs. $\mu$=4.00, $\sigma$=4.00; $Z$=-2.00; $p$<0.05).
For generation tasks, participants rated that they were significantly more satisfied with the final image ($\mu$=3.17, $\sigma$=3.00 vs. $\mu$=5.00, $\sigma$=5.50; $Z$=-2.17; $p$<0.05).
Significance was measured with the Wilcoxon Signed Rank test.

\ipstart{Gaining a summary of image content}
With \sysname{} across both tasks, all participants started by reading the summary table including the comparison description (summary of similarities and differences), as well as the per-image descriptions.  Participants all stated that the summary table was helpful for understanding the images they generated, as P6 explained: \textit{``I cannot do without the summary. Highlighting the differences was very useful.''} (P6). In addition, participants noted that the summary table's per-image descriptions were valuable for understanding the images. For example, P19 mentioned \textit{``This is more like an audio description because I can make a very clear mental image!''} and slowed down his screen reader pace to mimic the experience of listening to an audio description. P20 reported \textit{``I always thought that AI is not as capable of describing as humans, because usually alt-text generated by AI is short and doesn’t capture much information. But reading this, I am rethinking AI’s capabilities.''}. P12 found the detailed descriptions particularly helpful when authoring rather than interpreting images: \textit{``The first table (comparison description table) is so comprehensive. When I'm authoring images I need more information than when I'm looking at what others uploaded.''} (P12).

Using the baseline, participants all initially read all of the information they had access to (the caption and objects) for each image. all participants mentioned the inconvenience of having short image captions for gaining an overview, especially when the generated images are similar to each other. For example, after reading the BLIP-2 caption of S4, P18 asked \textit{``Are they all same images?''}

\ipstart{Selectively accessing additional information}
While all participants accessed the summary table first, we observed multiple strategies of using additional information provided by \sysname{} to understand the differences between the generated images. 
First, P9, P7, P16, P18, and P20 checked the information from all tables before making their decision. P20 mentioned \textit{``They are equally important but in different ways. If the generated images are different, the summary table would be sufficient. For similar ones, I’d have to go down the tables more.''} P16 noted \textit{``We never have too much information. All the details provided here matter to me''}.
After checking all the tables, P18 and P20 revisited the summary table again to remember and organize all information. 
The other seven participants (P10-P12, P8-P15, P17, P19) checked the tables selectively. Participants' preferences reflected their prior experiences creating images. For instance, P7 who typically creates images using an SVG editor prioritized the prompt verification table. He said ~\textit{``I detail more things in the prompt and want everything to be in the image, `cause I am more used to programming-drawing.''} P13 skipped the style and errors table as he was not familiar with the concepts despite the definitions provided: ~\textit{``As a born blind person, most information in the visual attributes is not useful as it's hard to imagine those.''} 
Participants also mentioned that they liked that \sysname{} provided the breakdown of the summary description into multiple tables. P16 described that \sysname{} has \textit{``So much transparency because it provides access to intermediate tables that constitute the summary table, just like a [prramming tool]! I can look at the inside of the models and see what they're doing.''} 
P10 and P11 both mentioned that they appreciated the order of the tables: \textit{``The summary [table] is the bigger picture. Then the tables go into the details. I also like that the prompt questions come first because they're important.''}

Participants also employed multiple strategies for navigating within the tables. 
Participants browsed through questions in the tables to identify questions they found to be important and skipped questions that were less important (\textit{e.g.}, not interested, or already appeared in the summary descriptions). 
We also identified multiple patterns of navigating within the tables. 
Participants checked all cells in a row when they found the table to be important. For instance, P11 checked the answers of all four images in the prompt verification table. In other cases, participants first checked the questions, then decided whether to read the row or skip to the next row. Participants skipped rows if the answers to the questions were already mentioned in the summary table, or if they were not interested in the question. For example, P8 skipped the medium, lighting, and perspective row in the visual style \& errors table and only attended to the error row. Sometimes, participants only checked the answer cells if the summary column highlighted the differences between the images and skipped to the next row if the summary stated mainly the similarities between the images.
Participants stated that \sysname{}'s table format was easy to navigate.  P19 noted the ease of navigation within the table: \textit{''I like having control with the tables. If the question or summary doesn't seem interesting, I can skip to the next row instead of reading all answers of four images.''}

\ipstart{Asking additional information}
With the baseline, most participants (12 participants in the interpretation task, 9 participants in the generation task) asked follow-up questions to try to understand the images, while with our system participants rarely asked follow-up questions (1 participant in the interpretation task and none in the generation task). P16 was the only participant who asked additional visual questions with \sysname{} after reading the table (\textit{`Is the data showed falling or rising?'} and \textit{`What is the date of the x-axis?'} for S3 in Figure~\ref{fig:materials}).
When asked about the reason for not asking any additional questions, P18 said \textit{``Looking at captions I already had a big picture so I didn’t ask additional questions.''} P7 similarly reflected: \textit{``I like that [\sysname] asks questions that I haven’t thought of but are still important. The answers to the questions told me additional stuff about the images.''} 
In contrast, with the baseline interface, participants asked many additional visual questions. Because each image was presented separately, participants often asked the same question for each image to compare the answers. Most of the questions were about the objects detected, especially when the object was not mentioned in the caption or did not seem relevant to the setting (\textit{e.g.,} P11 asked \textit{``Where is the beachball in the picture?''} after reading the object detection results of an image with the kitchen setting). P10 who experienced the baseline condition after \sysname{} reflected that \textit{``This one [Baseline] is not simply laid out for me. The previous one [\sysname{}] is easy peasy presenting everything for me. And this one is `Here you have to figure out.''}


\ipstart{Refining and Iterating Prompt}
In the generation task, none of the participants refined the prompt using the baseline and five participants refined the prompt when using \sysname{} (P9, P10, P13, P16, P17). Among the remaining 7 participants, 5 participants reported that they did not iterate as they were satisfied with the results, and 2 participants were unsure how to iterate the prompt after realizing that the image generation model did not reflect some parts of the original prompt (P15, P20). 

\begin{figure}[t]
  \centering
  \includegraphics[width=3.33in]{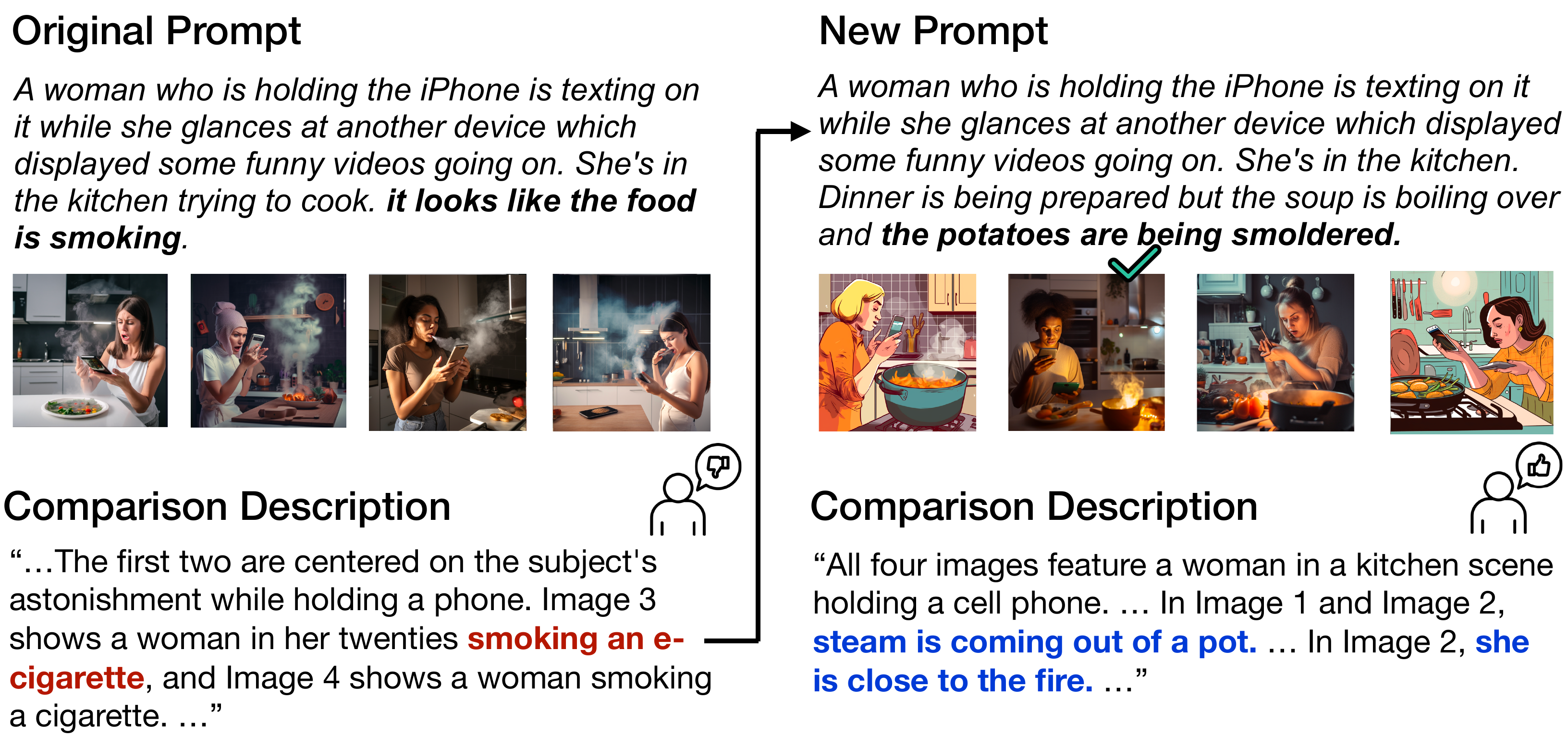}
  \caption{P10 generated the first set of images, noticed that the image generation model has made errors in the image (depicting the woman smoking instead of the food smoking), and corrected her prompt by replacing ``smoking'' with ``being smoldered''. }
  \label{fig:sample_iteration}
  \Description{This figure illustrates how P10 iterated the prompt after reading the comparison description. The original prompt was ``A woman who is holding the iPhone is texting on it while she glances at another device which displayed some funny videos going on. She's in the kitchen trying to cook. it looks like the food is smoking.'' and GenAssist generated the following comparison description: The first two are centered on the subject's astonishment while holding a phone. Image 3 shows a woman in her twenties smoking an e-cigarette, and Image 4 shows a woman smoking a cigarette. Then she came up with the new prompt ``A woman who is holding the iPhone is texting on it while she glances at another device which displayed some funny videos going on. She's in the kitchen. Dinner is being prepared but the soup is boiling over and the potatoes are being smoldered.'' and the new comparison description was All four images feature a woman in a kitchen scene holding a cell phone. ... In Image 1 and Image 2, steam is coming out of a pot. ... In Image 2, she is close to the fire.}
\end{figure}

Participants often quickly made the decision to revise the prompt while reading the summary table and before they moved on to other tables. For instance, while generating an image about an article about multitasking, P10 first attempted to generate an image with the following prompt \textit{`A woman who is holding the iPhone is texting on it while she glances at another device which displayed some funny videos going on. She's in the kitchen trying to cook. it looks like the food is smoking'}~\autoref{fig:sample_iteration}. However, she quickly noticed that most of the images generated depicted the woman as smoking instead of the food as smoking. She quickly iterated the prompt by replacing the word with \textit{`smoldering'} to generate a new set of images.

In addition, participants reported that \sysname{} informed them about the capabilities of the image generation model and guided them to refine their prompts. P20 mentioned \textit{``After reading the tables, it makes me think of what AI is capable of generating and what is not. It can't exactly reflect what I try to accomplish when the prompt is too complicated, so I will have to adjust my expectation and adjust my prompt.''} 
Participants also noted that \sysname{} is helpful for learning how to generate a detailed prompt (P7, P16, P17). P16 stated \textit{``Visual [styles \& errors] table is helpful for learning new styles.''} Similarly, P7 said \textit{``If I don't specify the styles, I think AI is generating [the styles] based on the context and content. So I know which style is good for which.''}

\ipstart{Selecting an Image Candidate}
To choose the final image from the four image candidates in the generation task, participants using \sysname{} often considered whether the image followed the prompt, whether additional details added by the generation model were relevant, and whether the image style or emotion was appropriate to the usage context. P17 said \textit{``I choose the third image because it has the information that I described.} Also, P7 mentioned \textit{``I will not choose the cartoon image because I want to be more serious here.''} Some participants changed the choice of image as they moved on to the next tables in the \sysname{}. For example, P8 who generated images of multiple plastic containers to portray the pollution problem updated his choice as he read the style and errors table: \textit{``Oh so the last image has many colors, I want to change to this one because I want it to be colorful!''} 

\ipstart{Noticed and unnoticed errors}
Participants encountered errors using both interfaces. In the baseline, all participants read the objects following the captions, but objects occasionally contained errors (\textit{e.g.,} labeling as another object that has similar shapes, colors, or textures). When the participants noticed objects irrelevant to the context, they often asked about the object but the questions about non-existent objects often led to further confusion. For instance, P11 asked \textit{`Where is the television?'} for an image where a television is not present. Because the answer generated by BLIP-2 was \textit{`There is no television.'}, P11 was more confused and did not consider the image due to uncertainty. Also, P16 asked \textit{`Where is the lollipop in the image?'} for an image without a lollipop (S1 in \autoref{fig:materials}) and BLIP-2 answered with a hallucination \textit{`In the man's mouth.'}, misleading P16. While \sysname{} features the same list of objects, participants did not experience this issue as they prioritized other information or recognized misinformation by referencing across multiple information sources. While using \sysname, P10, P7, P16 pointed out that some visual information in the tables conflicted with one another. For instance, in the second image of S2 (\autoref{fig:materials}), the summary table stated that the woman is walking in the street, but when the \sysname{} asked ~\textit{`Is she dancing?'} for prompt verification, BLIP-2 answered with \textit{`Yes'}, which confused the participants. P16 hypothesized that the caption mentioned walking because the dancing action is hard to capture in one image frame and thus the image is actually showing her dancing.
Still, participants did not notice inaccurate information in \sysname{} if there was no conflict. For example, a woman was described as looking happy but had a neutral expression (the 4th image for P10's 2nd prompt in~\autoref{fig:sample_iteration}). P10 removed the image from consideration as she wanted the woman to look stressed rather than happy.

\ipstart{Future improvements for \sysname{}}
Participants noted suggestions on how to improve \sysname{}'s description in the future.
First, P9 and P8 participants noted that the visual information provided by \sysname{} was long and difficult to process at once. This reflects users' subjective ratings on mental demand which is comparable in \sysname{} and the baseline in the interpretation task. Participants suggested allowing users to remove image columns and question rows from consideration. P8 mentioned \textit{``I want to filter images based on certain answers so that from then on, I won’t consider all four images and it will be easier!''}  P17 also shared that he wanted \sysname{} to learn from his interactions with the cells so that gradually it will present only the rows of interest.

Participants mentioned the difficulties of writing good prompts. P13 said \textit{``Even if I read the definitions about the style, it's hard to feel what effect it will give.''} In the generation task, none of the participants specified the medium in the prompt as they were not familiar with it. This often resulted in the image generations having varied styles. In addition, P7 and P16 mentioned that it is difficult to decide on what content to put in the prompt to effectively convey the message. P16 mentioned \textit{``I want to give it the whole book and make it generate.''} After experiencing that the generation model cannot reflect all the details in the prompt when the prompt is too long and complex, P12 stated \textit{``I want [\sysname{}] to tell me what [the generation model] can generate and what it can not.''}

\section{Discussion}
In this section, we reflect on our findings from the 
development and evaluation of \sysname. We also discuss future opportunities for research exploring accessible media authoring tools.

\ipstart{Scope of \sysname}
\revised{\sysname{} uses a text-to-image generation model~\cite{midjourney} to generate image candidates, vision-language models~\cite{radford2021learning, li2023blip} to extract visual information, and a large language model~\cite{openai2023gpt4} to synthesize descriptions. The scope of \sysname{} reflects the limitations of the models it uses. First, we designed \sysname{} to support the images that text-to-image generation models currently support: content-driven photos or illustrations with simple structures. However, both text-to-image generation and \sysname{} do not yet support images that are information-rich or densely structured such as information visualizations~\cite{sharif2022voxlens, sharif2023understanding} or diagrams~\cite{austin2023authoring, sorge2015end}. As text-to-image generation improves, future research will explore extending \sysname{} to complex graphics with text. For example, \sysname{} could help creators recognize if their prompt-generated diagram contains the desired text (by integrating Optical Character Recognition), relationships, and perceptual qualities (\textit{e.g.,} legibility, saliency of important information). 

Second, the descriptions that GenAssist is capable of providing are also limited by the capabilities of the pre-trained vision-language models~\cite{radford2021learning, li2023blip, zhou2022detecting}. For example, while GenAssist helped creators notice image generation errors such as omitted prompt details~\cite{liu2022design}, distortions to human bodies~\cite{wu2023better}, and objects placed illogically~\cite{xu2023imagereward}, some errors remained undetected. Also, \sysname{} occasionally included hallucinations (\textit{e.g.,} missing or non-existent objects) in the descriptions. While these issues may be mitigated with improvements to text-to-image models (\textit{e.g.,} better aligning with human preferences~\cite{wu2023better}) and vision language models (\textit{e.g.,} better composition reasoning~\cite{ma2023crepe}, reducing hallucinations~\cite{biten2022let}), \sysname{} could also learn what prompts are prone to generation errors and guide BLV creators in creating strong prompts.

Finally, while \sysname{}’s pipeline surfaced large differences between images (\textit{e.g.,} different objects, characters, expressions, or styles), its descriptions often missed smaller differences between images that were less likely to be described in training data captions (\textit{e.g.,} slightly different compositions or makeup styles). Thus, \sysname{} is currently useful in the early stages of prompt iteration, where large differences between images remain. In the future, \sysname{} could detect detailed changes by adding more detailed or domain-specific content and style questions, or integrating vision models that explicitly compare images~\cite{wang2020compare}.}

\ipstart{Understanding Multiple Images}
Creators in the formative study revealed that it is difficult to understand multiple images at the same time (\textbf{D2}. Understanding high-level image similarities and differences). 
To tackle this challenge, we designed \sysname{} with three strategies: (1) providing the overview of similarities and differences between the generated image candidates, (2) progressively disclosing the information from high-level to low-level to give the user control over the level of detail received~\cite{morris2018rich, huh2022cocomix, peng2023slidegestalt}, and (3) presenting the descriptions in a table format so that users can easily navigate between images to compare them. 
Participants highlighted that not only these detailed summaries but also the ability to selectively gain information about the underlying questions were helpful in narrowing down their choices. For example, some participants prioritized the prompt verification table to assess if the image followed their instructions (\textbf{D3}. Assessing if images followed the prompt), and other participants used the content and style table to learn how to improve their prompts (\textbf{D4}.  Accessing image details not specified by the prompt).
In the future, \sysname{} could support sorting or filtering images based on visual attributes to limit the number of images they consider at once (\textit{e.g., } sorting images based on prompt adherence or filtering images that have AI-generated distortions). \sysname{} could also read image descriptions with multiple voice styles to help creators distinguish generation candidates.

\sysname{}'s ability to attend to multiple similar images and surface differences can be useful in broader contexts. Our study participants expressed interest in using \sysname{} for comparing image search results or similar photos in social media. It can also help BLV people in decision-making situations based on visual information (\textit{e.g.,} online shopping, communicating with the design team in the software development, selecting a photo from similar shots).


\ipstart{Implications for Visual Question Answering} \revised{Comparing \sysname{} to our baseline of typical descriptions with visual question answering (VQA), all participants rated \sysname{} as more useful for understanding differences between images and creators asked fewer follow up questions with \sysname{}.
\sysname{} reduced follow-up questions by predicting visual questions based on the formative study and applying the questions to multiple images. Our \textit{predict-ask-summarize} approach also reduced the requirement for reading individual question answers. Future VQA systems intended for real-world environments may benefit from our approach as repetitive questions, ``unknown unknowns'', and complex visuals are likely.} 

\ipstart{Support in Creating Prompts}
In the formative study, we distilled the need to support creating prompts (\textbf{D1}. Authoring prompts that specify content and style). While we do not directly support prompt creation, we designed our system to reveal visual content and styles based on prompt guidelines to inform users about details the model filled in. In the user study, participants cited that reading the tables in \sysname{} helped inform their prompt iterations and learn about what styles to use.
Prior work has explored using structured search for visual concepts for writing prompts~\cite{liu2022opal, prompter}, and combining our system with such prior work is a promising avenue for future work. 
We are currently exploring suggesting content and styles for the prompt when the user specifies the context of image use and new ways to help users add specificity to their prompt (\textit{e.g.} a chatbot, as suggested in the formative study). 
In addition to text input, we can also consider multimodal input from users in the future such as image prompts~\cite{qiao2022initial}, sketch prompts~\cite{chen2009sketch2photo, zhang2023adding}, or music prompts~\cite{qiu2018image} to create an image for a music album cover, as desired by P6.

 \ipstart{Supporting Creators with Different Visual Impairments}
\revised{BLV creators’ interest in color or style information (\textit{e.g.}, medium, lighting, angle) often depended on their prior experience with visuals and onset of blindness. \sysname{} supports creators in selectively accessing description details, but in the future \sysname{} will let creators control which details to filter out or prioritize. To support creators without knowledge of visual style, \sysname{} could recommend popular styles given the image's intended use, provide style descriptions, or deliver style in another modality (\textit{e.g.,} sound~\cite{img-to-music}, tactile interfaces).
We will also improve GenAssist in the future to support users with remaining vision beyond providing descriptions. For example, GenAssist could provide descriptions based on the current zoom viewing window or support further visual edits to the generated images, as desired by P1.}


\revised{\ipstart{Implications of \sysname{} on Creativity}
Text-to-image generation models have sparked conversations about their implications for creativity. 
For BLV creators, image generation can improve creative agency compared to existing approaches for creating or selecting images. In our formative study, creators wanted to use image generation as it provided fewer limits over content and style than searching for images online and greater autonomy than asking a sighted person to create the image. \sysname{} supports BLV creators in exercising creative control over generated images by letting creators examine image details to revise the prompt or make an informed selection. Compared to sighted artists who use generated images primarily as references~\cite{liu2022opal}, BLV creators often intend to use generated images directly.
In the future, \sysname{} will further creative control by supporting prompt-based editing~\cite{bar2022text2live}.}

\ipstart{Implications of \sysname{} on Communication} 
We designed \sysname{} to support communication goals of BLV creators. BLV creators in our formative study aimed to create images to express their ideas to a broad audience and achieve self-expression. 
Images are particularly useful for capturing visual attention and communicating with sighted people who have difficulty reading text. 
For example, P4 generated an image of his family to share with his child. BLV creators also wanted to use \sysname{} in the workplace and on digital platforms. 
As \sysname{} exists in an ableist environment that prioritizes visual communication, there is a risk that \sysname{} may cause sighted people to expect image-based communication from BLV people. Tools like \sysname{} must be coupled with research and activism to make digital, workplace, and educational environments accessible --- \textit{e.g.,} enabling non-visual communication and providing access to existing visuals. Our work also reveals that generated images themselves should be shared with descriptions in addition to the prompt that might not accurately reflect the image. 


\ipstart{Generative AI for Accessible Media Authoring}
Advances in large-scale generative models enable people to create new types of content, yet no existing research has explored people with disabilities as the users of these tools~\cite{ko2023large}. 
We see opportunities for generative AI models to broaden the type of content that people with disabilities can create. For example, our study participants mentioned that they are interested in using generative models for creating dynamic graphics like cartoons and videos.
Similarly, generative models may be useful for people with motor impairments authoring visual media, or people with hearing impairments authoring music.



\section{Conclusion}
We created \sysname{}, an accessible text-to-image generation system for BLV creators. Informed by our formative study with 8 BLV creators, our interface enables users to verify the adherence of generated images to their prompts, access additional image details, and quickly assess similarities and differences between image candidates. Our system is powered by large language and vision-language models that generate visual questions, extract answers, and summarize the visual information. Our user study with 12 BLV creators demonstrated the effectiveness of our approach. We hope this research will catalyze future work in supporting people with disabilities to express their creativity.




\bibliographystyle{ACM-Reference-Format}
\bibliography{sample-base}


\begin{thebibliography}{85}


\ifx \showCODEN    \undefined \def \showCODEN     #1{\unskip}     \fi
\ifx \showDOI      \undefined \def \showDOI       #1{#1}\fi
\ifx \showISBNx    \undefined \def \showISBNx     #1{\unskip}     \fi
\ifx \showISBNxiii \undefined \def \showISBNxiii  #1{\unskip}     \fi
\ifx \showISSN     \undefined \def \showISSN      #1{\unskip}     \fi
\ifx \showLCCN     \undefined \def \showLCCN      #1{\unskip}     \fi
\ifx \shownote     \undefined \def \shownote      #1{#1}          \fi
\ifx \showarticletitle \undefined \def \showarticletitle #1{#1}   \fi
\ifx \showURL      \undefined \def \showURL       {\relax}        \fi
\providecommand\bibfield[2]{#2}
\providecommand\bibinfo[2]{#2}
\providecommand\natexlab[1]{#1}
\providecommand\showeprint[2][]{arXiv:#2}

\bibitem[Abid et~al\mbox{.}(2019)]%
        {abid2019gradio}
\bibfield{author}{\bibinfo{person}{Abubakar Abid}, \bibinfo{person}{Ali
  Abdalla}, \bibinfo{person}{Ali Abid}, \bibinfo{person}{Dawood Khan},
  \bibinfo{person}{Abdulrahman Alfozan}, {and} \bibinfo{person}{James Zou}.}
  \bibinfo{year}{2019}\natexlab{}.
\newblock \showarticletitle{Gradio: Hassle-free sharing and testing of ml
  models in the wild}.
\newblock \bibinfo{journal}{\emph{arXiv preprint arXiv:1906.02569}}
  (\bibinfo{year}{2019}).
\newblock


\bibitem[AccessiblePublishing.ca(2023)]%
        {image_desc_guideline}
\bibfield{author}{\bibinfo{person}{AccessiblePublishing.ca}.}
  \bibinfo{year}{2023 (accessed Apr 2, 2023)}\natexlab{}.
\newblock \bibinfo{title}{Guide to Image Descriptions}.
\newblock
\newblock
\urldef\tempurl%
\url{https://www.accessiblepublishing.ca/a-guide-to-image-description/}
\showURL{%
\tempurl}


\bibitem[Austin and Sorge(2023)]%
        {austin2023authoring}
\bibfield{author}{\bibinfo{person}{David Austin} {and} \bibinfo{person}{Volker
  Sorge}.} \bibinfo{year}{2023}\natexlab{}.
\newblock \showarticletitle{Authoring Web-accessible Mathematical Diagrams}. In
  \bibinfo{booktitle}{\emph{Proceedings of the 20th International Web for All
  Conference}}. \bibinfo{pages}{148--152}.
\newblock


\bibitem[Bar-Tal et~al\mbox{.}(2022)]%
        {bar2022text2live}
\bibfield{author}{\bibinfo{person}{Omer Bar-Tal}, \bibinfo{person}{Dolev
  Ofri-Amar}, \bibinfo{person}{Rafail Fridman}, \bibinfo{person}{Yoni Kasten},
  {and} \bibinfo{person}{Tali Dekel}.} \bibinfo{year}{2022}\natexlab{}.
\newblock \showarticletitle{Text2live: Text-driven layered image and video
  editing}. In \bibinfo{booktitle}{\emph{European conference on computer
  vision}}. Springer, \bibinfo{pages}{707--723}.
\newblock


\bibitem[Bennett et~al\mbox{.}(2018)]%
        {bennett2018teens}
\bibfield{author}{\bibinfo{person}{Cynthia~L Bennett}, \bibinfo{person}{Jane
  E}, \bibinfo{person}{Martez~E Mott}, \bibinfo{person}{Edward Cutrell}, {and}
  \bibinfo{person}{Meredith~Ringel Morris}.} \bibinfo{year}{2018}\natexlab{}.
\newblock \showarticletitle{How teens with visual impairments take, edit, and
  share photos on social media}. In \bibinfo{booktitle}{\emph{Proceedings of
  the 2018 CHI conference on human factors in computing systems}}.
  \bibinfo{pages}{1--12}.
\newblock


\bibitem[Bigham et~al\mbox{.}(2010)]%
        {bigham2010vizwiz}
\bibfield{author}{\bibinfo{person}{Jeffrey~P Bigham},
  \bibinfo{person}{Chandrika Jayant}, \bibinfo{person}{Hanjie Ji},
  \bibinfo{person}{Greg Little}, \bibinfo{person}{Andrew Miller},
  \bibinfo{person}{Robert~C Miller}, \bibinfo{person}{Robin Miller},
  \bibinfo{person}{Aubrey Tatarowicz}, \bibinfo{person}{Brandyn White},
  \bibinfo{person}{Samual White}, {et~al\mbox{.}}}
  \bibinfo{year}{2010}\natexlab{}.
\newblock \showarticletitle{Vizwiz: nearly real-time answers to visual
  questions}. In \bibinfo{booktitle}{\emph{Proceedings of the 23nd annual ACM
  symposium on User interface software and technology}}.
  \bibinfo{pages}{333--342}.
\newblock


\bibitem[Biten et~al\mbox{.}(2022)]%
        {biten2022let}
\bibfield{author}{\bibinfo{person}{Ali~Furkan Biten},
  \bibinfo{person}{Llu{\'\i}s G{\'o}mez}, {and} \bibinfo{person}{Dimosthenis
  Karatzas}.} \bibinfo{year}{2022}\natexlab{}.
\newblock \showarticletitle{Let there be a clock on the beach: Reducing object
  hallucination in image captioning}. In \bibinfo{booktitle}{\emph{Proceedings
  of the IEEE/CVF Winter Conference on Applications of Computer Vision}}.
  \bibinfo{pages}{1381--1390}.
\newblock


\bibitem[Bornschein and Weber(2017)]%
        {bornschein2017digital}
\bibfield{author}{\bibinfo{person}{Jens Bornschein} {and}
  \bibinfo{person}{Gerhard Weber}.} \bibinfo{year}{2017}\natexlab{}.
\newblock \showarticletitle{Digital drawing tools for blind users: A
  state-of-the-art and requirement analysis}. In
  \bibinfo{booktitle}{\emph{Proceedings of the 10th International Conference on
  Pervasive Technologies Related to Assistive Environments}}.
  \bibinfo{pages}{21--28}.
\newblock


\bibitem[Brady et~al\mbox{.}(2013)]%
        {brady2013visual}
\bibfield{author}{\bibinfo{person}{Erin Brady},
  \bibinfo{person}{Meredith~Ringel Morris}, \bibinfo{person}{Yu Zhong},
  \bibinfo{person}{Samuel White}, {and} \bibinfo{person}{Jeffrey~P Bigham}.}
  \bibinfo{year}{2013}\natexlab{}.
\newblock \showarticletitle{Visual challenges in the everyday lives of blind
  people}. In \bibinfo{booktitle}{\emph{Proceedings of the SIGCHI conference on
  human factors in computing systems}}. \bibinfo{pages}{2117--2126}.
\newblock


\bibitem[Brown et~al\mbox{.}(2020)]%
        {brown2020language}
\bibfield{author}{\bibinfo{person}{Tom Brown}, \bibinfo{person}{Benjamin Mann},
  \bibinfo{person}{Nick Ryder}, \bibinfo{person}{Melanie Subbiah},
  \bibinfo{person}{Jared~D Kaplan}, \bibinfo{person}{Prafulla Dhariwal},
  \bibinfo{person}{Arvind Neelakantan}, \bibinfo{person}{Pranav Shyam},
  \bibinfo{person}{Girish Sastry}, \bibinfo{person}{Amanda Askell},
  {et~al\mbox{.}}} \bibinfo{year}{2020}\natexlab{}.
\newblock \showarticletitle{Language models are few-shot learners}.
\newblock \bibinfo{journal}{\emph{Advances in neural information processing
  systems}}  \bibinfo{volume}{33} (\bibinfo{year}{2020}),
  \bibinfo{pages}{1877--1901}.
\newblock


\bibitem[Chen et~al\mbox{.}(2009)]%
        {chen2009sketch2photo}
\bibfield{author}{\bibinfo{person}{Tao Chen}, \bibinfo{person}{Ming-Ming
  Cheng}, \bibinfo{person}{Ping Tan}, \bibinfo{person}{Ariel Shamir}, {and}
  \bibinfo{person}{Shi-Min Hu}.} \bibinfo{year}{2009}\natexlab{}.
\newblock \showarticletitle{Sketch2photo: Internet image montage}.
\newblock \bibinfo{journal}{\emph{ACM transactions on graphics (TOG)}}
  \bibinfo{volume}{28}, \bibinfo{number}{5} (\bibinfo{year}{2009}),
  \bibinfo{pages}{1--10}.
\newblock


\bibitem[Devlin et~al\mbox{.}(2018)]%
        {devlin2018bert}
\bibfield{author}{\bibinfo{person}{Jacob Devlin}, \bibinfo{person}{Ming-Wei
  Chang}, \bibinfo{person}{Kenton Lee}, {and} \bibinfo{person}{Kristina
  Toutanova}.} \bibinfo{year}{2018}\natexlab{}.
\newblock \showarticletitle{Bert: Pre-training of deep bidirectional
  transformers for language understanding}.
\newblock \bibinfo{journal}{\emph{arXiv preprint arXiv:1810.04805}}
  (\bibinfo{year}{2018}).
\newblock


\bibitem[Dosovitskiy et~al\mbox{.}(2020)]%
        {dosovitskiy2020image}
\bibfield{author}{\bibinfo{person}{Alexey Dosovitskiy}, \bibinfo{person}{Lucas
  Beyer}, \bibinfo{person}{Alexander Kolesnikov}, \bibinfo{person}{Dirk
  Weissenborn}, \bibinfo{person}{Xiaohua Zhai}, \bibinfo{person}{Thomas
  Unterthiner}, \bibinfo{person}{Mostafa Dehghani}, \bibinfo{person}{Matthias
  Minderer}, \bibinfo{person}{Georg Heigold}, \bibinfo{person}{Sylvain Gelly},
  {et~al\mbox{.}}} \bibinfo{year}{2020}\natexlab{}.
\newblock \showarticletitle{An image is worth 16x16 words: Transformers for
  image recognition at scale}.
\newblock \bibinfo{journal}{\emph{arXiv preprint arXiv:2010.11929}}
  (\bibinfo{year}{2020}).
\newblock


\bibitem[Facebook(2021)]%
        {facebook-detailed-description}
\bibfield{author}{\bibinfo{person}{Facebook}.} \bibinfo{year}{2021}\natexlab{}.
\newblock \bibinfo{title}{How Facebook is using AI to improve photo
  descriptions for people who are blind or visually impaired}.
\newblock
\newblock
\urldef\tempurl%
\url{https://ai.facebook.com/blog/how-facebook-is-using-ai-to-improve-photo-descriptions-for-people-who-are-blind-or-visually-impaired/}
\showURL{%
\tempurl}


\bibitem[Falase et~al\mbox{.}(2019)]%
        {falase2019tactile}
\bibfield{author}{\bibinfo{person}{Olutayo Falase}, \bibinfo{person}{Alexa~F
  Siu}, {and} \bibinfo{person}{Sean Follmer}.} \bibinfo{year}{2019}\natexlab{}.
\newblock \showarticletitle{Tactile code skimmer: A tool to help blind
  programmers feel the structure of code}. In
  \bibinfo{booktitle}{\emph{Proceedings of the 21st International ACM SIGACCESS
  Conference on Computers and Accessibility}}. \bibinfo{pages}{536--538}.
\newblock


\bibitem[Goodfellow et~al\mbox{.}(2020)]%
        {goodfellow2020generative}
\bibfield{author}{\bibinfo{person}{Ian Goodfellow}, \bibinfo{person}{Jean
  Pouget-Abadie}, \bibinfo{person}{Mehdi Mirza}, \bibinfo{person}{Bing Xu},
  \bibinfo{person}{David Warde-Farley}, \bibinfo{person}{Sherjil Ozair},
  \bibinfo{person}{Aaron Courville}, {and} \bibinfo{person}{Yoshua Bengio}.}
  \bibinfo{year}{2020}\natexlab{}.
\newblock \showarticletitle{Generative adversarial networks}.
\newblock \bibinfo{journal}{\emph{Commun. ACM}} \bibinfo{volume}{63},
  \bibinfo{number}{11} (\bibinfo{year}{2020}), \bibinfo{pages}{139--144}.
\newblock


\bibitem[Gurari et~al\mbox{.}(2018)]%
        {gurari2018vizwiz}
\bibfield{author}{\bibinfo{person}{Danna Gurari}, \bibinfo{person}{Qing Li},
  \bibinfo{person}{Abigale~J Stangl}, \bibinfo{person}{Anhong Guo},
  \bibinfo{person}{Chi Lin}, \bibinfo{person}{Kristen Grauman},
  \bibinfo{person}{Jiebo Luo}, {and} \bibinfo{person}{Jeffrey~P Bigham}.}
  \bibinfo{year}{2018}\natexlab{}.
\newblock \showarticletitle{Vizwiz grand challenge: Answering visual questions
  from blind people}. In \bibinfo{booktitle}{\emph{Proceedings of the IEEE
  conference on computer vision and pattern recognition}}.
  \bibinfo{pages}{3608--3617}.
\newblock


\bibitem[https://github.com/mikhail bot/(2023)]%
        {stable_diffusion_negativeprompts}
\bibfield{author}{\bibinfo{person}{https://github.com/mikhail bot/}.}
  \bibinfo{year}{2023 (accessed Apr 2, 2023)}\natexlab{}.
\newblock \bibinfo{title}{Stable Diffusion Negative Prompts}.
\newblock
\newblock
\urldef\tempurl%
\url{https://github.com/mikhail-bot/stable-diffusion-negative-prompts}
\showURL{%
\tempurl}


\bibitem[https://github.com/pharmapsychotic/(2023)]%
        {interrogator}
\bibfield{author}{\bibinfo{person}{https://github.com/pharmapsychotic/}.}
  \bibinfo{year}{2023 (accessed Apr 2, 2023)}\natexlab{}.
\newblock \bibinfo{title}{CLIP Interrogator}.
\newblock
\newblock
\urldef\tempurl%
\url{https://github.com/pharmapsychotic/clip-interrogator}
\showURL{%
\tempurl}


\bibitem[https://github.com/willwulfken/(2023)]%
        {midjourney_styles}
\bibfield{author}{\bibinfo{person}{https://github.com/willwulfken/}.}
  \bibinfo{year}{2023 (accessed Apr 2, 2023)}\natexlab{}.
\newblock \bibinfo{title}{Midjourney Styles and Keywords}.
\newblock
\newblock
\urldef\tempurl%
\url{https://github.com/willwulfken/MidJourney-Styles-and-Keywords-Reference}
\showURL{%
\tempurl}


\bibitem[https://huggingface.co/spaces/fffiloni/(2023)]%
        {img-to-music}
\bibfield{author}{\bibinfo{person}{https://huggingface.co/spaces/fffiloni/}.}
  \bibinfo{year}{2023 (accessed Apr 2, 2023)}\natexlab{}.
\newblock \bibinfo{title}{Image to Music}.
\newblock
\newblock
\urldef\tempurl%
\url{https://huggingface.co/spaces/fffiloni/img-to-music}
\showURL{%
\tempurl}


\bibitem[Hu et~al\mbox{.}(2023)]%
        {hu2023tifa}
\bibfield{author}{\bibinfo{person}{Yushi Hu}, \bibinfo{person}{Benlin Liu},
  \bibinfo{person}{Jungo Kasai}, \bibinfo{person}{Yizhong Wang},
  \bibinfo{person}{Mari Ostendorf}, \bibinfo{person}{Ranjay Krishna}, {and}
  \bibinfo{person}{Noah~A Smith}.} \bibinfo{year}{2023}\natexlab{}.
\newblock \showarticletitle{TIFA: Accurate and Interpretable Text-to-Image
  Faithfulness Evaluation with Question Answering}.
\newblock \bibinfo{journal}{\emph{arXiv preprint arXiv:2303.11897}}
  (\bibinfo{year}{2023}).
\newblock


\bibitem[Huh et~al\mbox{.}(2022)]%
        {huh2022cocomix}
\bibfield{author}{\bibinfo{person}{Mina Huh}, \bibinfo{person}{YunJung Lee},
  \bibinfo{person}{Dasom Choi}, \bibinfo{person}{Haesoo Kim},
  \bibinfo{person}{Uran Oh}, {and} \bibinfo{person}{Juho Kim}.}
  \bibinfo{year}{2022}\natexlab{}.
\newblock \showarticletitle{Cocomix: Utilizing Comments to Improve Non-Visual
  Webtoon Accessibility}. In \bibinfo{booktitle}{\emph{Proceedings of the 2022
  CHI Conference on Human Factors in Computing Systems}}.
  \bibinfo{pages}{1--18}.
\newblock


\bibitem[Huh et~al\mbox{.}(2023)]%
        {huh2023avscript}
\bibfield{author}{\bibinfo{person}{Mina Huh}, \bibinfo{person}{Saelyne Yang},
  \bibinfo{person}{Yi-Hao Peng}, \bibinfo{person}{Xiang'Anthony' Chen},
  \bibinfo{person}{Young-Ho Kim}, {and} \bibinfo{person}{Amy Pavel}.}
  \bibinfo{year}{2023}\natexlab{}.
\newblock \showarticletitle{AVscript: Accessible Video Editing with
  Audio-Visual Scripts}. In \bibinfo{booktitle}{\emph{Proceedings of the 2023
  CHI Conference on Human Factors in Computing Systems}}.
\newblock


\bibitem[Kim et~al\mbox{.}(2023)]%
        {kim2023exploring}
\bibfield{author}{\bibinfo{person}{Jiho Kim}, \bibinfo{person}{Arjun
  Srinivasan}, \bibinfo{person}{Nam~Wook Kim}, {and} \bibinfo{person}{Yea-Seul
  Kim}.} \bibinfo{year}{CHI 2023}\natexlab{}.
\newblock \showarticletitle{Exploring Chart Question Answering for Blind and
  Low Vision Users}.
\newblock  (\bibinfo{year}{CHI 2023}).
\newblock


\bibitem[Kingma and Welling(2013)]%
        {kingma2013auto}
\bibfield{author}{\bibinfo{person}{Diederik~P Kingma} {and}
  \bibinfo{person}{Max Welling}.} \bibinfo{year}{2013}\natexlab{}.
\newblock \showarticletitle{Auto-encoding variational bayes}.
\newblock \bibinfo{journal}{\emph{arXiv preprint arXiv:1312.6114}}
  (\bibinfo{year}{2013}).
\newblock


\bibitem[Ko et~al\mbox{.}(2022)]%
        {ko2022we}
\bibfield{author}{\bibinfo{person}{Hyung-Kwon Ko}, \bibinfo{person}{Subin An},
  \bibinfo{person}{Gwanmo Park}, \bibinfo{person}{Seung~Kwon Kim},
  \bibinfo{person}{Daesik Kim}, \bibinfo{person}{Bohyoung Kim},
  \bibinfo{person}{Jaemin Jo}, {and} \bibinfo{person}{Jinwook Seo}.}
  \bibinfo{year}{2022}\natexlab{}.
\newblock \showarticletitle{We-toon: A Communication Support System between
  Writers and Artists in Collaborative Webtoon Sketch Revision}. In
  \bibinfo{booktitle}{\emph{Proceedings of the 35th Annual ACM Symposium on
  User Interface Software and Technology}}. \bibinfo{pages}{1--14}.
\newblock


\bibitem[Ko et~al\mbox{.}(2023)]%
        {ko2023large}
\bibfield{author}{\bibinfo{person}{Hyung-Kwon Ko}, \bibinfo{person}{Gwanmo
  Park}, \bibinfo{person}{Hyeon Jeon}, \bibinfo{person}{Jaemin Jo},
  \bibinfo{person}{Juho Kim}, {and} \bibinfo{person}{Jinwook Seo}.}
  \bibinfo{year}{2023}\natexlab{}.
\newblock \showarticletitle{Large-scale text-to-image generation models for
  visual artists’ creative works}. In \bibinfo{booktitle}{\emph{Proceedings
  of the 28th International Conference on Intelligent User Interfaces}}.
  \bibinfo{pages}{919--933}.
\newblock


\bibitem[Leake et~al\mbox{.}(2020)]%
        {leake2020generating}
\bibfield{author}{\bibinfo{person}{Mackenzie Leake},
  \bibinfo{person}{Hijung~Valentina Shin}, \bibinfo{person}{Joy~O Kim}, {and}
  \bibinfo{person}{Maneesh Agrawala}.} \bibinfo{year}{2020}\natexlab{}.
\newblock \showarticletitle{Generating Audio-Visual Slideshows from Text
  Articles Using Word Concreteness.}. In \bibinfo{booktitle}{\emph{CHI}},
  Vol.~\bibinfo{volume}{20}. \bibinfo{pages}{25--30}.
\newblock


\bibitem[Lee et~al\mbox{.}(2022b)]%
        {lee2022collabally}
\bibfield{author}{\bibinfo{person}{Cheuk Yin~Phipson Lee},
  \bibinfo{person}{Zhuohao Zhang}, \bibinfo{person}{Jaylin Herskovitz},
  \bibinfo{person}{JooYoung Seo}, {and} \bibinfo{person}{Anhong Guo}.}
  \bibinfo{year}{CHI 2022}\natexlab{b}.
\newblock \showarticletitle{CollabAlly: Accessible Collaboration Awareness in
  Document Editing}.
\newblock  (\bibinfo{year}{CHI 2022}).
\newblock


\bibitem[Lee et~al\mbox{.}(2022a)]%
        {lee2022imageexplorer}
\bibfield{author}{\bibinfo{person}{Jaewook Lee}, \bibinfo{person}{Jaylin
  Herskovitz}, \bibinfo{person}{Yi-Hao Peng}, {and} \bibinfo{person}{Anhong
  Guo}.} \bibinfo{year}{2022}\natexlab{a}.
\newblock \showarticletitle{ImageExplorer: Multi-Layered Touch Exploration to
  Encourage Skepticism Towards Imperfect AI-Generated Image Captions}. In
  \bibinfo{booktitle}{\emph{Proceedings of the 2022 CHI Conference on Human
  Factors in Computing Systems}}. \bibinfo{pages}{1--15}.
\newblock


\bibitem[Lee et~al\mbox{.}(2021)]%
        {lee2021image}
\bibfield{author}{\bibinfo{person}{Jaewook Lee}, \bibinfo{person}{Yi-Hao Peng},
  \bibinfo{person}{Jaylin Herskovitz}, {and} \bibinfo{person}{Anhong Guo}.}
  \bibinfo{year}{2021}\natexlab{}.
\newblock \showarticletitle{Image Explorer: Multi-Layered Touch Exploration to
  Make Images Accessible}. In \bibinfo{booktitle}{\emph{Proceedings of the 23rd
  International ACM SIGACCESS Conference on Computers and Accessibility}}.
  \bibinfo{pages}{1--4}.
\newblock


\bibitem[Li et~al\mbox{.}(2019)]%
        {li2019editing}
\bibfield{author}{\bibinfo{person}{Jingyi Li}, \bibinfo{person}{Son Kim},
  \bibinfo{person}{Joshua~A Miele}, \bibinfo{person}{Maneesh Agrawala}, {and}
  \bibinfo{person}{Sean Follmer}.} \bibinfo{year}{2019}\natexlab{}.
\newblock \showarticletitle{Editing spatial layouts through tactile templates
  for people with visual impairments}. In \bibinfo{booktitle}{\emph{Proceedings
  of the 2019 CHI Conference on Human Factors in Computing Systems}}.
  \bibinfo{pages}{1--11}.
\newblock


\bibitem[Li et~al\mbox{.}(2023)]%
        {li2023blip}
\bibfield{author}{\bibinfo{person}{Junnan Li}, \bibinfo{person}{Dongxu Li},
  \bibinfo{person}{Silvio Savarese}, {and} \bibinfo{person}{Steven Hoi}.}
  \bibinfo{year}{2023}\natexlab{}.
\newblock \showarticletitle{Blip-2: Bootstrapping language-image pre-training
  with frozen image encoders and large language models}.
\newblock \bibinfo{journal}{\emph{arXiv preprint arXiv:2301.12597}}
  (\bibinfo{year}{2023}).
\newblock


\bibitem[Li et~al\mbox{.}(2021)]%
        {li2021accessibility}
\bibfield{author}{\bibinfo{person}{Junchen Li}, \bibinfo{person}{Garreth
  W.~Tigwell}, {and} \bibinfo{person}{Kristen Shinohara}.}
  \bibinfo{year}{2021}\natexlab{}.
\newblock \showarticletitle{Accessibility of high-fidelity prototyping tools}.
  In \bibinfo{booktitle}{\emph{Proceedings of the 2021 CHI Conference on Human
  Factors in Computing Systems}}. \bibinfo{pages}{1--17}.
\newblock


\bibitem[Liu and Chilton(2022)]%
        {liu2022design}
\bibfield{author}{\bibinfo{person}{Vivian Liu} {and} \bibinfo{person}{Lydia~B
  Chilton}.} \bibinfo{year}{2022}\natexlab{}.
\newblock \showarticletitle{Design guidelines for prompt engineering
  text-to-image generative models}. In \bibinfo{booktitle}{\emph{Proceedings of
  the 2022 CHI Conference on Human Factors in Computing Systems}}.
  \bibinfo{pages}{1--23}.
\newblock


\bibitem[Liu et~al\mbox{.}(2022)]%
        {liu2022opal}
\bibfield{author}{\bibinfo{person}{Vivian Liu}, \bibinfo{person}{Han Qiao},
  {and} \bibinfo{person}{Lydia Chilton}.} \bibinfo{year}{2022}\natexlab{}.
\newblock \showarticletitle{Opal: Multimodal Image Generation for News
  Illustration}. In \bibinfo{booktitle}{\emph{Proceedings of the 35th Annual
  ACM Symposium on User Interface Software and Technology}}.
  \bibinfo{pages}{1--17}.
\newblock


\bibitem[Ma et~al\mbox{.}(2023)]%
        {ma2023crepe}
\bibfield{author}{\bibinfo{person}{Zixian Ma}, \bibinfo{person}{Jerry Hong},
  \bibinfo{person}{Mustafa~Omer Gul}, \bibinfo{person}{Mona Gandhi},
  \bibinfo{person}{Irena Gao}, {and} \bibinfo{person}{Ranjay Krishna}.}
  \bibinfo{year}{2023}\natexlab{}.
\newblock \showarticletitle{CREPE: Can Vision-Language Foundation Models Reason
  Compositionally?}. In \bibinfo{booktitle}{\emph{Proceedings of the IEEE/CVF
  Conference on Computer Vision and Pattern Recognition}}.
  \bibinfo{pages}{10910--10921}.
\newblock


\bibitem[McGeehan(2023)]%
        {prompter}
\bibfield{author}{\bibinfo{person}{Shane McGeehan}.} \bibinfo{year}{2023
  (accessed Apr 2, 2023)}\natexlab{}.
\newblock \bibinfo{title}{Prompter}.
\newblock
\newblock
\urldef\tempurl%
\url{https://prompterguide.com/prompter/}
\showURL{%
\tempurl}


\bibitem[Microsoft(2021)]%
        {seeingAI}
\bibfield{author}{\bibinfo{person}{Microsoft}.}
  \bibinfo{year}{2021}\natexlab{}.
\newblock \bibinfo{title}{Seeing AI}.
\newblock
\newblock
\urldef\tempurl%
\url{https://www.microsoft.com/en-us/ai/seeing-ai}
\showURL{%
\tempurl}


\bibitem[Midjourney(2023a)]%
        {midjourney}
\bibfield{author}{\bibinfo{person}{Midjourney}.} \bibinfo{year}{2023 (accessed
  Apr 2, 2023)}\natexlab{a}.
\newblock \bibinfo{title}{Midjourney}.
\newblock
\newblock
\urldef\tempurl%
\url{https://www.midjourney.com}
\showURL{%
\tempurl}


\bibitem[Midjourney(2023b)]%
        {midjourney_prompts}
\bibfield{author}{\bibinfo{person}{Midjourney}.} \bibinfo{year}{2023 (accessed
  Apr 2, 2023)}\natexlab{b}.
\newblock \bibinfo{title}{Midjourney Propmt Guidelines}.
\newblock
\newblock
\urldef\tempurl%
\url{https://docs.midjourney.com/docs/prompts}
\showURL{%
\tempurl}


\bibitem[Morris et~al\mbox{.}(2018)]%
        {morris2018rich}
\bibfield{author}{\bibinfo{person}{Meredith~Ringel Morris},
  \bibinfo{person}{Jazette Johnson}, \bibinfo{person}{Cynthia~L Bennett}, {and}
  \bibinfo{person}{Edward Cutrell}.} \bibinfo{year}{2018}\natexlab{}.
\newblock \showarticletitle{Rich representations of visual content for screen
  reader users}. In \bibinfo{booktitle}{\emph{Proceedings of the 2018 CHI
  conference on human factors in computing systems}}. \bibinfo{pages}{1--11}.
\newblock


\bibitem[News(2016)]%
        {formative_article}
\bibfield{author}{\bibinfo{person}{Hospital News}.}
  \bibinfo{year}{2016}\natexlab{}.
\newblock \bibinfo{title}{You are what you eat}.
\newblock
\newblock
\urldef\tempurl%
\url{https://hospitalnews.com/you-are-what-you-eat-why-nutrition-matters/}
\showURL{%
\tempurl}


\bibitem[O'Donovan et~al\mbox{.}(2015)]%
        {o2015designscape}
\bibfield{author}{\bibinfo{person}{Peter O'Donovan}, \bibinfo{person}{Aseem
  Agarwala}, {and} \bibinfo{person}{Aaron Hertzmann}.}
  \bibinfo{year}{2015}\natexlab{}.
\newblock \showarticletitle{Designscape: Design with interactive layout
  suggestions}. In \bibinfo{booktitle}{\emph{Proceedings of the 33rd annual ACM
  conference on human factors in computing systems}}.
  \bibinfo{pages}{1221--1224}.
\newblock


\bibitem[OpenAI(2023)]%
        {openai2023gpt4}
\bibfield{author}{\bibinfo{person}{OpenAI}.} \bibinfo{year}{2023}\natexlab{}.
\newblock \bibinfo{title}{GPT-4 Technical Report}.
\newblock
\newblock
\showeprint[arxiv]{2303.08774}~[cs.CL]


\bibitem[Parsons(2023)]%
        {dalle_promptbook}
\bibfield{author}{\bibinfo{person}{Guy Parsons}.} \bibinfo{year}{2023 (accessed
  Apr 2, 2023)}\natexlab{}.
\newblock \bibinfo{title}{DALL-E2 Propmt Book}.
\newblock
\newblock
\urldef\tempurl%
\url{https://dallery.gallery/the-dalle-2-prompt-book/}
\showURL{%
\tempurl}


\bibitem[Payne et~al\mbox{.}(2020)]%
        {payne2020blind}
\bibfield{author}{\bibinfo{person}{William~Christopher Payne},
  \bibinfo{person}{Alex~Yixuan Xu}, \bibinfo{person}{Fabiha Ahmed},
  \bibinfo{person}{Lisa Ye}, {and} \bibinfo{person}{Amy Hurst}.}
  \bibinfo{year}{2020}\natexlab{}.
\newblock \showarticletitle{How blind and visually impaired composers,
  producers, and songwriters leverage and adapt music technology}. In
  \bibinfo{booktitle}{\emph{Proceedings of the 22nd International ACM SIGACCESS
  Conference on Computers and Accessibility}}. \bibinfo{pages}{1--12}.
\newblock


\bibitem[Peng et~al\mbox{.}(2021a)]%
        {peng2021slidecho}
\bibfield{author}{\bibinfo{person}{Yi-Hao Peng}, \bibinfo{person}{Jeffrey~P
  Bigham}, {and} \bibinfo{person}{Amy Pavel}.}
  \bibinfo{year}{2021}\natexlab{a}.
\newblock \showarticletitle{Slidecho: Flexible Non-Visual Exploration of
  Presentation Videos}. In \bibinfo{booktitle}{\emph{The 23rd International ACM
  SIGACCESS Conference on Computers and Accessibility}}.
  \bibinfo{pages}{1--12}.
\newblock


\bibitem[Peng et~al\mbox{.}(2023)]%
        {peng2023slidegestalt}
\bibfield{author}{\bibinfo{person}{Yi-Hao Peng}, \bibinfo{person}{Peggy Chi},
  \bibinfo{person}{Anjuli Kannan}, \bibinfo{person}{Meredith Morris}, {and}
  \bibinfo{person}{Irfan Essa}.} \bibinfo{year}{2023}\natexlab{}.
\newblock \showarticletitle{Slide Gestalt: Automatic Structure Extraction in
  Slide Decks for Non-Visual Access}.
\newblock  (\bibinfo{year}{2023}).
\newblock


\bibitem[Peng et~al\mbox{.}(2021b)]%
        {peng2021say}
\bibfield{author}{\bibinfo{person}{Yi-Hao Peng}, \bibinfo{person}{JiWoong
  Jang}, \bibinfo{person}{Jeffrey~P Bigham}, {and} \bibinfo{person}{Amy
  Pavel}.} \bibinfo{year}{2021}\natexlab{b}.
\newblock \showarticletitle{Say It All: Feedback for Improving Non-Visual
  Presentation Accessibility}. In \bibinfo{booktitle}{\emph{Proceedings of the
  2021 CHI Conference on Human Factors in Computing Systems}}.
  \bibinfo{pages}{1--12}.
\newblock


\bibitem[Peng et~al\mbox{.}(2022)]%
        {peng2022diffscriber}
\bibfield{author}{\bibinfo{person}{Yi-Hao Peng}, \bibinfo{person}{Jason Wu},
  \bibinfo{person}{Jeffrey Bigham}, {and} \bibinfo{person}{Amy Pavel}.}
  \bibinfo{year}{2022}\natexlab{}.
\newblock \showarticletitle{Diffscriber: Describing Visual Design Changes to
  Support Mixed-Ability Collaborative Presentation Authoring}. In
  \bibinfo{booktitle}{\emph{Proceedings of the 35th Annual ACM Symposium on
  User Interface Software and Technology}}. \bibinfo{pages}{1--13}.
\newblock


\bibitem[Potluri et~al\mbox{.}(2019)]%
        {potluri2019multi}
\bibfield{author}{\bibinfo{person}{Venkatesh Potluri}, \bibinfo{person}{Liang
  He}, \bibinfo{person}{Christine Chen}, \bibinfo{person}{Jon~E Froehlich},
  {and} \bibinfo{person}{Jennifer Mankoff}.} \bibinfo{year}{2019}\natexlab{}.
\newblock \showarticletitle{A multi-modal approach for blind and visually
  impaired developers to edit webpage designs}. In
  \bibinfo{booktitle}{\emph{Proceedings of the 21st International ACM SIGACCESS
  Conference on Computers and Accessibility}}. \bibinfo{pages}{612--614}.
\newblock


\bibitem[Qiao et~al\mbox{.}(2022)]%
        {qiao2022initial}
\bibfield{author}{\bibinfo{person}{Han Qiao}, \bibinfo{person}{Vivian Liu},
  {and} \bibinfo{person}{Lydia Chilton}.} \bibinfo{year}{2022}\natexlab{}.
\newblock \showarticletitle{Initial Images: Using Image Prompts to Improve
  Subject Representation in Multimodal AI Generated Art}. In
  \bibinfo{booktitle}{\emph{Creativity and Cognition}}.
  \bibinfo{pages}{15--28}.
\newblock


\bibitem[Qiu and Kataoka(2018)]%
        {qiu2018image}
\bibfield{author}{\bibinfo{person}{Yue Qiu} {and} \bibinfo{person}{Hirokatsu
  Kataoka}.} \bibinfo{year}{2018}\natexlab{}.
\newblock \showarticletitle{Image generation associated with music data}. In
  \bibinfo{booktitle}{\emph{Proceedings of the IEEE Conference on Computer
  Vision and Pattern Recognition Workshops}}. \bibinfo{pages}{2510--2513}.
\newblock


\bibitem[Radford et~al\mbox{.}(2021)]%
        {radford2021learning}
\bibfield{author}{\bibinfo{person}{Alec Radford}, \bibinfo{person}{Jong~Wook
  Kim}, \bibinfo{person}{Chris Hallacy}, \bibinfo{person}{Aditya Ramesh},
  \bibinfo{person}{Gabriel Goh}, \bibinfo{person}{Sandhini Agarwal},
  \bibinfo{person}{Girish Sastry}, \bibinfo{person}{Amanda Askell},
  \bibinfo{person}{Pamela Mishkin}, \bibinfo{person}{Jack Clark},
  {et~al\mbox{.}}} \bibinfo{year}{2021}\natexlab{}.
\newblock \showarticletitle{Learning transferable visual models from natural
  language supervision}. In \bibinfo{booktitle}{\emph{International conference
  on machine learning}}. PMLR, \bibinfo{pages}{8748--8763}.
\newblock


\bibitem[Ramesh et~al\mbox{.}(2022)]%
        {ramesh2022hierarchical}
\bibfield{author}{\bibinfo{person}{Aditya Ramesh}, \bibinfo{person}{Prafulla
  Dhariwal}, \bibinfo{person}{Alex Nichol}, \bibinfo{person}{Casey Chu}, {and}
  \bibinfo{person}{Mark Chen}.} \bibinfo{year}{2022}\natexlab{}.
\newblock \showarticletitle{Hierarchical text-conditional image generation with
  clip latents}.
\newblock \bibinfo{journal}{\emph{arXiv preprint arXiv:2204.06125}}
  (\bibinfo{year}{2022}).
\newblock


\bibitem[Ramesh et~al\mbox{.}(2021)]%
        {ramesh2021zero}
\bibfield{author}{\bibinfo{person}{Aditya Ramesh}, \bibinfo{person}{Mikhail
  Pavlov}, \bibinfo{person}{Gabriel Goh}, \bibinfo{person}{Scott Gray},
  \bibinfo{person}{Chelsea Voss}, \bibinfo{person}{Alec Radford},
  \bibinfo{person}{Mark Chen}, {and} \bibinfo{person}{Ilya Sutskever}.}
  \bibinfo{year}{2021}\natexlab{}.
\newblock \showarticletitle{Zero-shot text-to-image generation}. In
  \bibinfo{booktitle}{\emph{International Conference on Machine Learning}}.
  PMLR, \bibinfo{pages}{8821--8831}.
\newblock


\bibitem[Reddy et~al\mbox{.}(2021)]%
        {reddy2021dall}
\bibfield{author}{\bibinfo{person}{Mr~D~Murahari Reddy},
  \bibinfo{person}{Mr~Sk~Masthan Basha}, \bibinfo{person}{Mr~M~Chinnaiahgari
  Hari}, {and} \bibinfo{person}{Mr~N Penchalaiah}.}
  \bibinfo{year}{2021}\natexlab{}.
\newblock \showarticletitle{Dall-e: Creating images from text}.
\newblock \bibinfo{journal}{\emph{UGC Care Group I Journal}}
  \bibinfo{volume}{8}, \bibinfo{number}{14} (\bibinfo{year}{2021}),
  \bibinfo{pages}{71--75}.
\newblock


\bibitem[Rombach et~al\mbox{.}(2021)]%
        {rombach2021highresolution}
\bibfield{author}{\bibinfo{person}{Robin Rombach}, \bibinfo{person}{Andreas
  Blattmann}, \bibinfo{person}{Dominik Lorenz}, \bibinfo{person}{Patrick
  Esser}, {and} \bibinfo{person}{Björn Ommer}.}
  \bibinfo{year}{2021}\natexlab{}.
\newblock \bibinfo{title}{High-Resolution Image Synthesis with Latent Diffusion
  Models}.
\newblock
\newblock
\showeprint[arxiv]{2112.10752}~[cs.CV]


\bibitem[Schaadhardt et~al\mbox{.}(2021)]%
        {schaadhardt2021understanding}
\bibfield{author}{\bibinfo{person}{Anastasia Schaadhardt},
  \bibinfo{person}{Alexis Hiniker}, {and} \bibinfo{person}{Jacob~O Wobbrock}.}
  \bibinfo{year}{2021}\natexlab{}.
\newblock \showarticletitle{Understanding blind screen-reader users’
  experiences of digital artboards}. In \bibinfo{booktitle}{\emph{Proceedings
  of the 2021 CHI Conference on Human Factors in Computing Systems}}.
  \bibinfo{pages}{1--19}.
\newblock


\bibitem[Schuhmann et~al\mbox{.}(2022)]%
        {schuhmann2022laion}
\bibfield{author}{\bibinfo{person}{Christoph Schuhmann},
  \bibinfo{person}{Romain Beaumont}, \bibinfo{person}{Richard Vencu},
  \bibinfo{person}{Cade Gordon}, \bibinfo{person}{Ross Wightman},
  \bibinfo{person}{Mehdi Cherti}, \bibinfo{person}{Theo Coombes},
  \bibinfo{person}{Aarush Katta}, \bibinfo{person}{Clayton Mullis},
  \bibinfo{person}{Mitchell Wortsman}, {et~al\mbox{.}}}
  \bibinfo{year}{2022}\natexlab{}.
\newblock \showarticletitle{Laion-5b: An open large-scale dataset for training
  next generation image-text models}.
\newblock \bibinfo{journal}{\emph{arXiv preprint arXiv:2210.08402}}
  (\bibinfo{year}{2022}).
\newblock


\bibitem[Sefid et~al\mbox{.}(2021)]%
        {sefid2021slidegen}
\bibfield{author}{\bibinfo{person}{Athar Sefid}, \bibinfo{person}{Prasenjit
  Mitra}, {and} \bibinfo{person}{Lee Giles}.} \bibinfo{year}{2021}\natexlab{}.
\newblock \showarticletitle{SlideGen: an abstractive section-based slide
  generator for scholarly documents}. In \bibinfo{booktitle}{\emph{Proceedings
  of the 21st ACM Symposium on Document Engineering}}. \bibinfo{pages}{1--4}.
\newblock


\bibitem[Sharif et~al\mbox{.}(2022)]%
        {sharif2022voxlens}
\bibfield{author}{\bibinfo{person}{Ather Sharif}, \bibinfo{person}{Olivia~H
  Wang}, \bibinfo{person}{Alida~T Muongchan}, \bibinfo{person}{Katharina
  Reinecke}, {and} \bibinfo{person}{Jacob~O Wobbrock}.}
  \bibinfo{year}{2022}\natexlab{}.
\newblock \showarticletitle{Voxlens: Making online data visualizations
  accessible with an interactive javascript plug-in}. In
  \bibinfo{booktitle}{\emph{Proceedings of the 2022 CHI Conference on Human
  Factors in Computing Systems}}. \bibinfo{pages}{1--19}.
\newblock


\bibitem[Sharif et~al\mbox{.}(2023)]%
        {sharif2023understanding}
\bibfield{author}{\bibinfo{person}{Ather Sharif}, \bibinfo{person}{Andrew~M
  Zhang}, \bibinfo{person}{Katharina Reinecke}, {and} \bibinfo{person}{Jacob~O
  Wobbrock}.} \bibinfo{year}{2023}\natexlab{}.
\newblock \showarticletitle{Understanding and Improving Drilled-Down
  Information Extraction from Online Data Visualizations for Screen-Reader
  Users}. In \bibinfo{booktitle}{\emph{Proceedings of the 20th International
  Web for All Conference}}. \bibinfo{pages}{18--31}.
\newblock


\bibitem[Sorge et~al\mbox{.}(2015)]%
        {sorge2015end}
\bibfield{author}{\bibinfo{person}{Volker Sorge}, \bibinfo{person}{Mark Lee},
  {and} \bibinfo{person}{Sandy Wilkinson}.} \bibinfo{year}{2015}\natexlab{}.
\newblock \showarticletitle{End-to-end solution for accessible chemical
  diagrams}. In \bibinfo{booktitle}{\emph{Proceedings of the 12th International
  Web for All Conference}}. \bibinfo{pages}{1--10}.
\newblock


\bibitem[Times(2023a)]%
        {plastic}
\bibfield{author}{\bibinfo{person}{NY Times}.} \bibinfo{year}{2023 (accessed
  Apr 2, 2023)}\natexlab{a}.
\newblock \bibinfo{title}{My Kids Want Plastic Toys. I Want to Go Green.}
\newblock
\newblock
\urldef\tempurl%
\url{https://time.com/6126981/my-kids-want-plastic-toys-i-want-to-go-green-heres-a-fix/}
\showURL{%
\tempurl}


\bibitem[Times(2023b)]%
        {multitasking}
\bibfield{author}{\bibinfo{person}{NY Times}.} \bibinfo{year}{2023 (accessed
  Apr 2, 2023)}\natexlab{b}.
\newblock \bibinfo{title}{Why Multitasking is Bad for You}.
\newblock
\newblock
\urldef\tempurl%
\url{https://time.com/4737286/multitasking-mental-health-stress-texting-depression/}
\showURL{%
\tempurl}


\bibitem[Turc and Nemade(2022)]%
        {midjourney_dataset}
\bibfield{author}{\bibinfo{person}{Iulia Turc} {and} \bibinfo{person}{Gaurav
  Nemade}.} \bibinfo{year}{2022}\natexlab{}.
\newblock \bibinfo{title}{Midjourney User Prompts \& Generated Images (250k)}.
\newblock
\newblock
\urldef\tempurl%
\url{https://doi.org/10.34740/KAGGLE/DS/2349267}
\showDOI{\tempurl}


\bibitem[Vaswani et~al\mbox{.}(2017)]%
        {vaswani2017attention}
\bibfield{author}{\bibinfo{person}{Ashish Vaswani}, \bibinfo{person}{Noam
  Shazeer}, \bibinfo{person}{Niki Parmar}, \bibinfo{person}{Jakob Uszkoreit},
  \bibinfo{person}{Llion Jones}, \bibinfo{person}{Aidan~N Gomez},
  \bibinfo{person}{{\L}ukasz Kaiser}, {and} \bibinfo{person}{Illia
  Polosukhin}.} \bibinfo{year}{2017}\natexlab{}.
\newblock \showarticletitle{Attention is all you need}.
\newblock \bibinfo{journal}{\emph{Advances in neural information processing
  systems}}  \bibinfo{volume}{30} (\bibinfo{year}{2017}).
\newblock


\bibitem[Vinyals et~al\mbox{.}(2015)]%
        {vinyals2015show}
\bibfield{author}{\bibinfo{person}{Oriol Vinyals}, \bibinfo{person}{Alexander
  Toshev}, \bibinfo{person}{Samy Bengio}, {and} \bibinfo{person}{Dumitru
  Erhan}.} \bibinfo{year}{2015}\natexlab{}.
\newblock \showarticletitle{Show and tell: A neural image caption generator}.
  In \bibinfo{booktitle}{\emph{Proceedings of the IEEE conference on computer
  vision and pattern recognition}}. \bibinfo{pages}{3156--3164}.
\newblock


\bibitem[Von~Ahn and Dabbish(2004)]%
        {von2004labeling}
\bibfield{author}{\bibinfo{person}{Luis Von~Ahn} {and} \bibinfo{person}{Laura
  Dabbish}.} \bibinfo{year}{2004}\natexlab{}.
\newblock \showarticletitle{Labeling images with a computer game}. In
  \bibinfo{booktitle}{\emph{Proceedings of the SIGCHI conference on Human
  factors in computing systems}}. \bibinfo{pages}{319--326}.
\newblock


\bibitem[(WAI)(2022)]%
        {wai}
\bibfield{author}{\bibinfo{person}{W3C Web Accessibility~Initiative (WAI)}.}
  \bibinfo{year}{2022 (accessed Dec 12, 2022)}\natexlab{}.
\newblock \bibinfo{title}{Introduction to web accessibility}.
\newblock
\newblock
\urldef\tempurl%
\url{https://www.w3.org/WAI/fundamentals/accessibility-intro/}
\showURL{%
\tempurl}


\bibitem[Wang et~al\mbox{.}(2020)]%
        {wang2020compare}
\bibfield{author}{\bibinfo{person}{Jiuniu Wang}, \bibinfo{person}{Wenjia Xu},
  \bibinfo{person}{Qingzhong Wang}, {and} \bibinfo{person}{Antoni~B Chan}.}
  \bibinfo{year}{2020}\natexlab{}.
\newblock \showarticletitle{Compare and reweight: Distinctive image captioning
  using similar images sets}. In \bibinfo{booktitle}{\emph{Computer
  Vision--ECCV 2020: 16th European Conference, Glasgow, UK, August 23--28,
  2020, Proceedings, Part I 16}}. Springer, \bibinfo{pages}{370--386}.
\newblock


\bibitem[Wang et~al\mbox{.}(2021)]%
        {wang2021revamp}
\bibfield{author}{\bibinfo{person}{Ruolin Wang}, \bibinfo{person}{Zixuan Chen},
  \bibinfo{person}{Mingrui~Ray Zhang}, \bibinfo{person}{Zhaoheng Li},
  \bibinfo{person}{Zhixiu Liu}, \bibinfo{person}{Zihan Dang},
  \bibinfo{person}{Chun Yu}, {and} \bibinfo{person}{Xiang'Anthony' Chen}.}
  \bibinfo{year}{2021}\natexlab{}.
\newblock \showarticletitle{Revamp: Enhancing Accessible Information Seeking
  Experience of Online Shopping for Blind or Low Vision Users}. In
  \bibinfo{booktitle}{\emph{Proceedings of the 2021 CHI Conference on Human
  Factors in Computing Systems}}. \bibinfo{pages}{1--14}.
\newblock


\bibitem[Wang et~al\mbox{.}(2023)]%
        {wang2023reprompt}
\bibfield{author}{\bibinfo{person}{Yunlong Wang}, \bibinfo{person}{Shuyuan
  Shen}, {and} \bibinfo{person}{Brian~Y Lim}.} \bibinfo{year}{2023}\natexlab{}.
\newblock \showarticletitle{RePrompt: Automatic Prompt Editing to Refine
  AI-Generative Art Towards Precise Expressions}.
\newblock \bibinfo{journal}{\emph{arXiv preprint arXiv:2302.09466}}
  (\bibinfo{year}{2023}).
\newblock


\bibitem[Wu et~al\mbox{.}(2017)]%
        {wu2017automatic}
\bibfield{author}{\bibinfo{person}{Shaomei Wu}, \bibinfo{person}{Jeffrey
  Wieland}, \bibinfo{person}{Omid Farivar}, {and} \bibinfo{person}{Julie
  Schiller}.} \bibinfo{year}{2017}\natexlab{}.
\newblock \showarticletitle{Automatic alt-text: Computer-generated image
  descriptions for blind users on a social network service}. In
  \bibinfo{booktitle}{\emph{proceedings of the 2017 ACM conference on computer
  supported cooperative work and social computing}}.
  \bibinfo{pages}{1180--1192}.
\newblock


\bibitem[Wu et~al\mbox{.}(2023)]%
        {wu2023better}
\bibfield{author}{\bibinfo{person}{Xiaoshi Wu}, \bibinfo{person}{Keqiang Sun},
  \bibinfo{person}{Feng Zhu}, \bibinfo{person}{Rui Zhao}, {and}
  \bibinfo{person}{Hongsheng Li}.} \bibinfo{year}{2023}\natexlab{}.
\newblock \showarticletitle{Better aligning text-to-image models with human
  preference}.
\newblock \bibinfo{journal}{\emph{arXiv preprint arXiv:2303.14420}}
  (\bibinfo{year}{2023}).
\newblock


\bibitem[Xia(2020)]%
        {xia2020crosspower}
\bibfield{author}{\bibinfo{person}{Haijun Xia}.}
  \bibinfo{year}{2020}\natexlab{}.
\newblock \showarticletitle{Crosspower: Bridging graphics and linguistics}. In
  \bibinfo{booktitle}{\emph{Proceedings of the 33rd Annual ACM Symposium on
  User Interface Software and Technology}}. \bibinfo{pages}{722--734}.
\newblock


\bibitem[Xu et~al\mbox{.}(2023)]%
        {xu2023imagereward}
\bibfield{author}{\bibinfo{person}{Jiazheng Xu}, \bibinfo{person}{Xiao Liu},
  \bibinfo{person}{Yuchen Wu}, \bibinfo{person}{Yuxuan Tong},
  \bibinfo{person}{Qinkai Li}, \bibinfo{person}{Ming Ding},
  \bibinfo{person}{Jie Tang}, {and} \bibinfo{person}{Yuxiao Dong}.}
  \bibinfo{year}{2023}\natexlab{}.
\newblock \showarticletitle{Imagereward: Learning and evaluating human
  preferences for text-to-image generation}.
\newblock \bibinfo{journal}{\emph{arXiv preprint arXiv:2304.05977}}
  (\bibinfo{year}{2023}).
\newblock


\bibitem[Xu et~al\mbox{.}(2015)]%
        {xu2015show}
\bibfield{author}{\bibinfo{person}{Kelvin Xu}, \bibinfo{person}{Jimmy Ba},
  \bibinfo{person}{Ryan Kiros}, \bibinfo{person}{Kyunghyun Cho},
  \bibinfo{person}{Aaron Courville}, \bibinfo{person}{Ruslan Salakhudinov},
  \bibinfo{person}{Rich Zemel}, {and} \bibinfo{person}{Yoshua Bengio}.}
  \bibinfo{year}{2015}\natexlab{}.
\newblock \showarticletitle{Show, attend and tell: Neural image caption
  generation with visual attention}. In \bibinfo{booktitle}{\emph{International
  conference on machine learning}}. PMLR, \bibinfo{pages}{2048--2057}.
\newblock


\bibitem[Zhang and Agrawala(2023)]%
        {zhang2023adding}
\bibfield{author}{\bibinfo{person}{Lvmin Zhang} {and} \bibinfo{person}{Maneesh
  Agrawala}.} \bibinfo{year}{2023}\natexlab{}.
\newblock \bibinfo{title}{Adding Conditional Control to Text-to-Image Diffusion
  Models}.
\newblock
\newblock
\showeprint[arxiv]{2302.05543}~[cs.CV]


\bibitem[Zhang and Wobbrock(2023)]%
        {zhang2023a11yboard}
\bibfield{author}{\bibinfo{person}{Zhuohao Zhang} {and}
  \bibinfo{person}{Jacob~O. Wobbrock}.} \bibinfo{year}{CHI 2023}\natexlab{}.
\newblock \showarticletitle{A11yBoard: Making Digital Artboards Accessible to
  Blind and Low-Vision Users}.
\newblock


\bibitem[Zhong et~al\mbox{.}(2015)]%
        {zhong2015regionspeak}
\bibfield{author}{\bibinfo{person}{Yu Zhong}, \bibinfo{person}{Walter~S
  Lasecki}, \bibinfo{person}{Erin Brady}, {and} \bibinfo{person}{Jeffrey~P
  Bigham}.} \bibinfo{year}{2015}\natexlab{}.
\newblock \showarticletitle{Regionspeak: Quick comprehensive spatial
  descriptions of complex images for blind users}. In
  \bibinfo{booktitle}{\emph{Proceedings of the 33rd Annual ACM Conference on
  Human Factors in Computing Systems}}. \bibinfo{pages}{2353--2362}.
\newblock


\bibitem[Zhou et~al\mbox{.}(2022)]%
        {zhou2022detecting}
\bibfield{author}{\bibinfo{person}{Xingyi Zhou}, \bibinfo{person}{Rohit
  Girdhar}, \bibinfo{person}{Armand Joulin}, \bibinfo{person}{Philipp
  Kr{\"a}henb{\"u}hl}, {and} \bibinfo{person}{Ishan Misra}.}
  \bibinfo{year}{2022}\natexlab{}.
\newblock \showarticletitle{Detecting Twenty-thousand Classes using Image-level
  Supervision}. In \bibinfo{booktitle}{\emph{ECCV}}.
\newblock


\end{thebibliography}


\clearpage
\onecolumn
\appendix
\section{STUDY PARTICIPANTS DEMOGRAPHICS}

\begin{table*}[h]
\small\sffamily\def\arraystretch{1.2}\setlength{\tabcolsep}{0.5em}
    \centering
    \begin{tabular}{lllllll}
        \toprule
       PID  & Gender & Age & Visual Impairment & Onset & Job & Images Produced \\
       \midrule
        P1 & Non-binary & 40 & Legally blind & Congenital & Artist & Paintings, Cartoons \\
        P2 & Male & 50 & Totally blind & Congenital & Professor (CS) & Presentations, Scientific figures \\
        P3 & Female & 29 & Legally blind & Congenital & Teacher (English) & Presentations, Course website \\
        P4 & Male & 28 & Totally blind & Acquired & Teacher (Music) & Website logos \\
        P5 & Male & 59 & Totally blind & Congenital & Professor (Climate) & Presentations, Scientific figures \\
        P6 & Male & 42 & Totally blind & Acquired & Software engineer & Website images, Music album cover \\
        P7 & Male & 32 & Totally blind & Acquired & Software engineer & Website images \\
        P8 & Male & 30 & Totally blind & Acquired & Graduate student & Presentations \\
        P9 & Female & 41 & Totally blind & Congenital & Graduate student & Presentations, Social media images \\
        P10 & Female & 30 & Totally blind & Acquired & Graduate student & Presentations, Website images \\
        P11 & Female & 37 & Totally blind & Congenital & Accessibility consultant & Website images \\
        P12 & Male & 50 & Legally blind & Totally blind & Finance consultant & Charts, Graphs \\        
        P13 & Male & 61 & Totally blind & Congenital & YouTuber, Musician & Video thumbnails \\
        P14 & Male & 44 & Totally blind & Congenital & Author, Photographer & Book covers \\
        P15 & Male & 20 & Totally blind & Congenital & University student & Book covers \\
        P16 & Male & 36 & Totally blind & Congenital & Artist & Event flyers \\
        P17 & Male & 26 & Totally blind & Congenital & Accessibility consultant & Icons, Video thumbnails \\
        P18 & Male & 47 & Legally blind & Acquired & Software engineer & Brochures, Website images \\
        
        \bottomrule
    \end{tabular}
    \caption{Participant table for formative and comparison study.}
    \label{tab:form_participants}
\end{table*}

\end{document}